\documentclass[12pt,reprint,aps,groupedaddress,showpacs,nofootinbib,prd]{revtex4-1}

\usepackage{amsmath,amssymb,graphicx,amsfonts}
\usepackage{mathrsfs}
\usepackage{comment}
\usepackage{xcolor}
\usepackage[shortlabels]{enumitem}
\usepackage{hyperref}
\usepackage{url}
\usepackage[title]{appendix}

\newcommand{\beq}{\begin{equation}}
	\newcommand{\eeq}{\end{equation}}
\newcommand{\bea}{\begin{eqnarray}}
	\newcommand{\eea}{\end{eqnarray}}
\newcommand{\bit}{\begin{itemize}}
	\newcommand{\eit}{\end{itemize}}
\newcommand{\ben}{\begin{enumerate}}
	\newcommand{\een}{\end{enumerate}}
\newcommand{\nn}{\nonumber}

\def\scri{\mathscr{I}}

\begin{document}
\title{Pseudospectrum of Reissner-Nordstr\"om black holes: \\
quasinormal mode instability and universality}

\author{Kyriakos Destounis$^{1,2}$, Rodrigo Panosso Macedo$^{2,3}$, Emanuele Berti$^4$, Vitor Cardoso$^2$, Jos\'e Luis Jaramillo$^5$} 
\affiliation{$^1$Theoretical Astrophysics, IAAT, University of T{\"u}bingen, 72076 T{\"u}bingen, Germany}
\affiliation{${^2}$CENTRA, Departamento de F\'{\i}sica, Instituto Superior T\'ecnico -- IST, Universidade de Lisboa -- UL, Avenida Rovisco Pais 1, 1049 Lisboa, Portugal}
\affiliation{$^3$School of Mathematical Sciences, Queen Mary, University of
	London, Mile End Road, London E1 4NS, United Kingdom}
\affiliation{$^4$Department of Physics and Astronomy, Johns Hopkins University, 3400 N. Charles Street, Baltimore, Maryland, 21218, USA}
\affiliation{$^5$Institut de Math\'ematiques de Bourgogne (IMB), UMR 5584, CNRS, Universit\'e de Bourgogne Franche-Comt\'e, F-21000 Dijon, France}

\begin{abstract}
Black hole spectroscopy is a powerful tool to probe the Kerr nature of  astrophysical compact objects and their environment. The observation of multiple ringdown modes in gravitational waveforms could soon lead to high-precision gravitational spectroscopy, so it is critical to understand if the quasinormal mode spectrum is stable against perturbations. It was recently shown that the pseudospectrum can shed light on the spectral stability of black hole quasinormal modes. We study the pseudospectrum of Reissner-Nordstr\"om spacetimes and we find a spectral instability of scalar and gravitoelectric quasinormal modes in subextremal and extremal black holes, extending similar findings for the Schwarzschild spacetime. The asymptotic structure of pseudospectral contour levels is the same for scalar and gravitoelectric perturbations. By making different gauge choices in the hyperboloidal slicing of the spacetime, we find that the broad features of the pseudospectra are remarkably gauge-independent. The gravitational-led and electromagnetic-led quasinormal modes of extremal Reissner-Nordstr\"om black holes exhibit ``strong" isospectrality: not only their spectrum coincides, but the whole pseudospectrum is the same for both classes of perturbations. We observe that a conformal duality between the extremal horizon and spacetime boundaries at infinity is responsible for such ``strong" isospectrality property.
\end{abstract}
	
\maketitle

\section{Introduction}

The observation of gravitational waves (GWs) from compact binaries has become a primary avenue of scientific exploration. A plethora of events have flooded ground-based detectors, leading to the emergence of a novel era of GW astronomy and black hole (BH) spectroscopy. Systematic GW observations from LIGO and Virgo~\cite{LIGOScientific:2018mvr,Abbott:2020niy}, as well as future ground- and space-based interferometers~\cite{LISA:2017pwj,TianQin:2015yph,Hu:2017mde}, will improve our understanding of gravity in the strong-field regime~\cite{Berti:2015itd,Barack:2018yly,Perkins:2020tra}. 

GWs carry pristine information on strong-field gravity, and in particular on compact objects and their environment. According to uniqueness theorems in general relativity, the merger of two isolated BHs eventually leads to a stationary BH described by at most three parameters: mass, charge (which is astrophysically expected to be negligible), and angular momentum~\cite{Robinson:1975bv,Bekenstein:1996pn,Chrusciel:2012jk,Cardoso:2016ryw}. This state is approached via a characteristic relaxation stage (the ``ringdown'') of the final distorted BH: the GW signal after coalescence is well described by a superposition of exponentially damped sinusoids. 
The oscillation frequencies and decay time scales form a discrete set of complex numbers, the so-called quasinormal mode (QNM) frequencies, which contain specific information on the underlying geometry~\cite{Kokkotas:1999bd,Berti:2009kk,Konoplya:2011qq}. The relaxation (a consequence of the dissipative nature of GWs) implies, mathematically, that QNMs are (generically) not a complete set. The timescales involved are similar to the energy levels of atoms and molecules, and they can reveal the structure of the compact object producing the radiation~\cite{Chandrasekhar:1975zza,Leaver:1986gd,Cardoso:2008bp}.

The QNM spectrum of BHs in general relativity is well understood~\cite{Kokkotas:1999bd,Berti:2009kk,Konoplya:2011qq,Berti:2005ys}, while the QNM {\it content} of the signal generated by the coalescence of compact objects (i.e., the relative amplitudes and phases of the modes) is less understood, but there are good indications that several modes -- including higher overtones and different multipolar components -- are important to fully understand the signal~\cite{Leaver:1986gd,Berti:2006wq,Berti:2005ys,Baibhav:2017jhs}. The analysis of recent GW events shows evidence for more than one mode, even at relatively low signal-to-noise ratios~\cite{Isi:2019aib,Capano:2021etf}. Upgrades to the existing GW facilities should lead to routine detections of BH ringdown signals with higher signal-to-noise ratios, heralding the new field of BH spectroscopy~\cite{Berti:2005ys,Berti:2007zu,Berti:2016lat,Cardoso:2017cqb,Berti:2018vdi,Cardoso:2019rvt}.

\subsection{Black hole quasinormal mode instability}

To fully exploit the potential of BH spectroscopy we must better understand the relative excitation of each QNM~\cite{Nollert:1998ys}, the sensitivity of the BH response to fluctuations {\it close} to these resonances, and the possible instability of the spectrum itself under small perturbations of the scattering potentials. Because astrophysical BHs are not isolated, this last question is of paramount importance. Early investigations found that BH QNMs are exponentially sensitive to small perturbations, either due to (far-away or nearby) matter~\cite{Nollert:1996rf,Nollert:1998ys,Leung:1999rh,Leung:1999iq,Barausse:2014tra} or due to variations in the boundary conditions~\cite{Cardoso:2016rao,Cardoso:2016oxy,Cardoso:2019rvt}. The pioneering work by Nollert and Price~\cite{Nollert:1996rf,Nollert:1998ys} has been recently extended to more general types of ``ultraviolet'' (small-scale) perturbations~\cite{Daghigh:2020jyk,Jaramillo:2020tuu,Qian:2020cnz,Liu:2021aqh,Jaramillo:2021tmt}.
Spectral instabilities are common to other dissipative systems. A spectral analysis may be insufficient to understand the response of systems affected by such instabilities, requiring the development of alternative tools~\cite{Trefethen:2005,Sjostrand2019}.

Spectral instabilities can have important implications for physical systems. This is well illustrated by the case of hydrodynamics, where theoretical predictions of the onset of turbulent flow based on eigenvalue analyses agree poorly with experiments~\cite{Trefethen:1993}. Similarly, the introduction of non-Hermitian (non-selfadjoint) operators in PT-symmetric quantum mechanics entails that the associated spectra contain insufficient information to draw full, quantum-mechanically relevant conclusions~\cite{Krejcirik:2014kaa}. Most importantly, one-dimensional wave equations with dissipative boundary conditions -- analogous to those that apply to perturbed BHs in spherical symmetry -- suffer from similar limitations in their spectral predictions~\cite{Driscoll:1996}. The common feature among these different physical problems is their formulation in terms of non-selfadjoint operators. 

For selfadjoint operators, the spectral theorem underlies the notion of normal modes (which provide an orthonormal basis) and guarantees the stability of the eigenvalues under perturbations. In other words, a small-scale perturbation to the operator leads to spectral values migrating in the complex plane within a region of size comparable to the scale of the perturbation. In stark contrast, the lack of such a theorem in the non-selfadjoint case entails, in general, the loss of completeness in the set of eigenfunctions as well as their orthogonality, possibly leading to spectral instabilities. Thus, the eigenvalues may show a strong sensitivity to small-scale perturbations. In these cases, the spectral points migrate to an extent that is orders of magnitude larger than the perturbation scale. This feature of non-selfadjoint operators (more generally, non-normal operators) is called spectral instability, and it is related to the loss of collinearity of ``left'' (bra) and ``right'' (ket) eigenvectors corresponding to a given eigenvalue.

\subsection{Pseudospectrum and universality of quasinormal modes}
The pseudospectrum is the formal mathematical concept capturing the extent to which systems controlled by non-selfadjoint operators exhibit spectral instabilities. Pseudospectral contour levels portray a ``topographical map'' of spectral migration, identifying the region in the complex plane where QNMs can migrate. Of particular importance is the behavior of pseudospectral contour levels at large real values of the QNM frequency: these are intimately related with the notion of QNM-free regions, namely the regions in the complex plane to where QNMs cannot migrate. QNM-free regions of general scatterers are known to belong to ``universal'' classes \cite{zworski2017mathematical}, whose parameters are controlled by the qualitative properties of the underlying system.

In the context of BH perturbation theory, Ref.~\cite{Jaramillo:2020tuu} presented a systematic framework to address BH QNM instability based on the notion of pseudospectrum, performing a comprehensive study of the Schwarzschild case. These results indicate a direct connection between the pseudospectral contour lines of the unperturbed Schwarzschild potential and the open branches (called ``Nollert-Price'' QNM branches in~\cite{Jaramillo:2020tuu})
formed by migrating perturbed QNM overtones, which resemble the w-mode spectra of neutron stars~\cite{Kokkotas:1992xak,ZhaWuLeu11}. Although QNMs are in principle ``free'' to move above the QNM-free regions bounded by the pseudospectral lines, the results in Refs.~\cite{Jaramillo:2020tuu,Jaramillo:2021tmt} show that QNM frequencies typically approach pseudospectral contours for perturbations of sufficiently large wave number (i.e., probing small scales) in patterns that seem independent of the detailed nature of such ultraviolet perturbations. Besides being useful to assess the spectral instability of BH QNMs, the universal asymptotics of pseudospectral contour lines are then good indicators of perturbed QNM branches, and therefore they hint at a possible universality in the asymptotics of QNM spectra of generic compact objects~\cite{Jaramillo:2020tuu}.

Another important property of BH QNM spectra is isospectrality. This property is a delicate feature of specific BH spacetimes~\cite{Chandrasekhar:1985kt,Nichols:2012jn,Cardoso:2019mqo,Moulin:2019bfh} and it is absent, for example, in compact stars. A better understanding of isospectrality breaking can offer hints of possible universal features in the QNM spectra of compact astrophysical objects. For Schwarzschild BHs, Ref.~\cite{Jaramillo:2021tmt} shows the existence of different regimes of isospectrality loss in different types of perturbed QNM branches. A systematic interpolation between BHs and compact stars -- in particular in terms of inner boundary conditions: see e.g. the work of Ref.~\cite{Maggio:2020jml}, based on the membrane paradigm -- can be used to improve our understanding of QNM isospectrality.

\subsection{The Reissner-Nordstr\"om spacetime}

In this paper we study the pseudospectrum of Reissner-Nordstr\"om (RN) BHs with mass $M$ and charge $Q$. As the closest non-trivial extension of Schwarzschild, the RN spacetime provides a well-controlled model to systematically extend the exploration of universality properties of BH pseudospectra initiated in \cite{Jaramillo:2020tuu,Jaramillo:2021tmt}, as well as inquiring into the proposed possible connection with the QNMs of generic compact objects. Indeed, the possibility of varying a parameter in a whole family of potentials permits to test the universality hypotheses in a well-defined setting: if BH QNM universality, controlled by asymptotically similar pseudospectra, is violated in this simple case, this probably rules out universality in realistic settings where matter plays a role. 

From a technical perspective, the RN solution allows us to test a geometrical aspect of universality, namely its ``spacetime slicing'' independence. In our approach, the calculation of BH pseudospectra relies on the so-called hyperboloidal framework~\cite{Zenginoglu:2007jw,Zenginoglu:2011jz,Ansorg:2016ztf,PanossoMacedo:2018hab,PanossoMacedo:2018gvw,PanossoMacedo:2019npm,Jaramillo:2020tuu}, where the dissipative boundary conditions at the BH horizon and in the wave zone are geometrically incorporated into the problem via a choice of constant-time slices intersecting future null infinity $\scri^+$ and the BH horizon ${\cal H}^+$. The hyperboloidal framework can be implemented using different slices, raising the question of the possible (gauge) dependence of the pseudospectrum on the adopted coordinates. The explicit construction of two independent coordinate systems for the RN spacetime~\cite{PanossoMacedo:2018hab}, reviewed in Sec.~\ref{section RN Hyperboloidal}, shows that different gauges yield consistent results, giving strong support to the geometrical nature of the pseudospectrum.

Finally, RN has features that are absent in the Schwarzschild case, such as the appearance of a family of near-extremal, long-lived modes in the extremal limit $Q\to M$~\cite{Kim:2012mh,Zimmerman:2015trm,Richartz:2015saa,Cardoso:2017soq}. These zero-quality factor modes can dominate the BH response to perturbations.
Extremal RN geometries are marginally stable under neutral massless scalar perturbations~\cite{Aretakis:2011ha,Aretakis:2011hc} and they can develop local horizon hair~\cite{Angelopoulos:2018yvt}. Moreover, gravitational-led QNMs with angular index $\ell$ coincide with electromagnetic-led QNMs with angular index $\ell-1$ in the extremal limit~\cite{Onozawa:1996ba,Okamura:1997ic,Kallosh:1997ug,Berti:2004md}, providing an intriguing testing ground for pseudospectral calculations that probe the near-resonance region. We will show that this symmetry is not broken away from the resonances.

\section{Reissner-Nordstr\"om perturbations in the hyperboloidal framework}

We are interested in static, spherically symmetric spacetimes described by the line element
\begin{equation}
ds^2 = -f(r) dt^2 + f(r)^{-1} dr^2 + r^2 \left({d \theta^2} + \sin^2\theta d\varphi^2\right),\label{line element}
\end{equation}
where $t=\text{constant}$ slices correspond to Cauchy surfaces which intersect the horizon bifurcation sphere and spatial infinity $i^0$.  For charged BH spacetimes, described by the RN geometry, the (square of the) lapse function $f(r)$ is
\begin{align}
f(r)= 1-\dfrac{2M}{r} + \dfrac{Q^2}{r^2} = \left( 1 - \dfrac{r_+}{r}\right)\left( 1 - \dfrac{r_-}{r}\right),\label{lapse}
\end{align}
where $M$ and $Q$ are the BH mass and electric charge, while $r=r_-$ and $r=r_+$ are the Cauchy and event horizon radii, respectively, such that $f(r_\pm)=0$. The horizons of RN geometries are explicitly given by
\begin{equation}
\label{RN horizons}
r_{\pm} =  M \pm \sqrt{M^2 - Q^2}.
\end{equation}
It is convenient to introduce a tortoise coordinate such that $r_*\in ]-\infty, +\infty[$ and $dr_*/dr=1/f(r)$. Explicitly, the tortoise coordinate reads
\beq
\label{tortoise RN}
\dfrac{r_*}{r_+} = \dfrac{r}{r_+} + \dfrac{1}{1-\kappa^2}\bigg[ \ln\left( \dfrac{r}{r_+} - 1 \right) - \kappa^4 \ln\left( \dfrac{r}{r_+} - \kappa^2 \right) \bigg],
\eeq
where\footnote{Note that Ref.~\cite{PanossoMacedo:2018hab} used a different definition for $\kappa$. One must replace $\kappa \rightarrow \kappa^2$ when comparing expressions from Ref.~\cite{PanossoMacedo:2018hab} with the ones presented here.  } $\kappa \equiv {Q}/{r_+} \in [-1,1]$. In particular, the asymptotic regions in the BH exterior correspond to $r=r_+$ ($r_*\rightarrow -\infty$) and $r\rightarrow +\infty$ ($r_*\rightarrow +\infty$).

From Eq.~\eqref{RN horizons} we have $Q^2=r_+ r_-$, $2M = r_+ + r_-$, and therefore
\begin{equation}
\kappa^2=\dfrac{r_-}{r_+}, \quad \dfrac{M}{r_+} = \dfrac{1+\kappa^2}{2}.
\end{equation}
We can express $\kappa$ in terms of the more common dimensionless charge parameter $Q/M$ as
\beq
\kappa = \dfrac{{Q/M}}{1+\sqrt{1-(Q/M)^2}}.
\eeq

\subsection{Perturbations of charged black holes}

The dynamics of scalar, electromagnetic and gravitational fields in the RN background is described by a second-order partial differential (wave) equation of the form
\begin{equation}\label{evolution equation}
\left(\frac{\partial^2}{\partial t^2} -\frac{\partial^2}{\partial r_*^2} + V \right)\phi = 0.
\end{equation}
Here, $\phi$ is a master wavefunction, which is a combination of the fundamental perturbed quantities. The effective potential $V$ depends on
the nature of the field. We focus here on scalar fields, and on (polar) gravitoelectric fluctuations. The effective potential for such perturbations can be written in the compact form~\cite{Moncrief74a,Moncrief74b,Moncrief75,Chandrasekhar:1985kt,Berti:2009kk,PanossoMacedo:2018hab} 
\begin{equation}
V= \dfrac{f(r)}{r^2}\left[ \ell(\ell+1) + \dfrac{r_+}{r}\left(\mu - \kappa^2 \nu \dfrac{r_+}{r}\right)  \right],\label{eq:potential_RN}
\end{equation}
with
\begin{align}\label{potential factors}
\nu &= 3 n_p - 1, \quad \mu = n_p(1+\kappa^2) - (1-n_p)\mathfrak{m}_\pm, \nonumber\\
\mathfrak{m}_\pm &= \dfrac{1+\kappa^2}{4} \Bigg[1 \pm 3 \sqrt{1 + \dfrac{4\kappa^2 A}{(1+\kappa^2)^2}}\Bigg], \\
A &= \dfrac{4}{9}(\ell+2)(\ell-1),\nonumber
\end{align}
where $\ell$ is the angular index of the perturbation. In the above parametrization, scalar perturbations are recovered for $n_p=1$, whereas electromagnetic-led $(\mathfrak{m}_-)$ and gravitational-led $(\mathfrak{m}_+)$ perturbations (which reduce to electromagnetic and gravitational perturbations of Schwarzschild BHs in the uncharged limit) are recovered for $n_p=-1$. 

\subsection{The hyperboloidal framework}\label{section RN Hyperboloidal}

To study BH resonances and pseudospectra, we adopt an approach in which the relevant wave-like operators are considered on a compact spatial domain. The hyperboloidal approach provides a geometric framework to compactify the wave equation along spatial directions and, in particular, study QNMs. The advantage of this scheme lies in the fact that the outgoing boundary conditions at the event horizon and infinity, which are fundamental for dissipative systems like BHs, are geometrically imposed by shifting the Cauchy slice $\Sigma_t$ appropriately so that it intersects future null infinity $\scri^+$ and the BH event horizon ${\cal H}^+$. In what follows, we summarize the basic ingredients of this framework (see also Refs.~\cite{Zenginoglu:2007jw,Zenginoglu:2011jz,Ansorg:2016ztf,PanossoMacedo:2018hab,PanossoMacedo:2018gvw,PanossoMacedo:2019npm,Jaramillo:2020tuu} and references therein; in particular, Ref.~\cite{PanossoMacedo:2018hab} gives more details on the hyperboloidal framework in the RN spacetime employed in this work).

A practical way to introduce the hyperboloidal approach follows from the coordinate transformation
\begin{equation}
\begin{aligned}
\dfrac{t}{\lambda} = \tau - h(\sigma), \quad 
\dfrac{r_*}{\lambda} = g(\sigma),
\end{aligned}\label{transformation}
\end{equation}
with $\lambda$ an appropriate length scale. The so-called height function~\cite{Zenginoglu:2007jw,Zenginoglu:2011jz} $h(\sigma)$ ``bends'' the original Cauchy slice $\Sigma_t$ so that $\tau=\text{constant}$ corresponds to hypersurfaces $\Sigma_\tau$ which penetrate the BH horizon and intersect null infinity. The function $g(\sigma)$ introduces a spatial compactification from $r_*\in\left]-\infty,\infty\right[$ to a bounded interval $\sigma\in\left[0,1\right]$. The wave zone is now explicitly included in the domain, with $\sigma=0$ and $\sigma=1$ representing future null infinity and the BH horizon, respectively.

Upon the hyperboloidal coordinate transformation \eqref{transformation}, a conformal rescaling of the line element \eqref{line element} can be performed via $d\tilde s^2 = \Omega^{2} ds^2$, with the conformal factor being directly associated to the radial coordinate $\sigma$ via $\Omega = \sigma/\lambda$. Future null infinity is then characterized by $\left.\Omega\right|_{\scri^+}=0$, whereas the conformal metric $d\tilde s^2$ is regular in the entire domain $\sigma\in[0,1]$. Ref.~\cite{PanossoMacedo:2018hab} provides explicit expressions for the conformal RN metric in the so-called minimal gauge. We will review this gauge in the next sections with focus, however, on the wave equation dictating the dynamics of scalar and gravitoelectric fluctuations propagating on this spacetime.

\subsubsection{Wave equation}

Under the coordinate change~\eqref{transformation}, the wave equation \eqref{evolution equation} acquires the form
\begin{equation}\label{hyperboloidal evolution equation}
- \ddot \phi + L_1 \phi + L_2 \dot \phi = 0,
\end{equation}
with $\dot{} = \partial_\tau$, and the differential operators given by~\cite{Jaramillo:2020tuu}
\begin{align}\label{L1}
L_1 &= \frac{1}{w(\sigma)}\big[\partial_\sigma\left(p(\sigma)\partial_\sigma\right) - q_\ell(\sigma)\big], \\\label{L2}
L_2 &= \frac{1}{w(\sigma)}\big[2\gamma(\sigma)\partial_\sigma + \partial_\sigma\gamma(\sigma)\big].
\end{align}
The height and compactification functions $h(\sigma)$ and $g(\sigma)$ enter the above expression via
\begin{equation}\label{new operators}
\begin{aligned}
& w(\sigma)=\frac{g'^2-h'^2}{|g'|}, \ \ p(\sigma) = |g'|^{-1},  \\
& \gamma(\sigma)=\frac{h'}{|g'|}, \ \ \ \ \ \ \ \ \ q(\sigma)= \lambda^2 |g'|\;V. 
\end{aligned}
\end{equation}
Fundamental for the hyperboloidal approach is the fact that $p(\sigma)$ vanishes at the domain boundaries $\sigma=0$ and $\sigma=1$. This property implies that the (Sturm-Liouville) operator $L_1$ is singular. Thus, the physically relevant solutions (describing ingoing waves at the horizon and outgoing waves at future null infinity) are those satisfying the equation's underlying regularity conditions.

The advantageous structure of Eq.~\eqref{hyperboloidal evolution equation} becomes obvious when one performs a first-order reduction in time, to rewrite it as a matrix evolution problem. By introducing $\psi=\dot \phi$, Eq.~\eqref{hyperboloidal evolution equation} reads
\begin{equation}\label{matrix evolution}
\dot u=i L u, \quad 	L =\frac{1}{i}\!
\left(
\begin{array}{c  c}
	0 & 1 \\
	L_1 & L_2
\end{array}
\right), \quad u=\left(
\begin{array}{c}\phi \\ \psi \end{array}\right),
\end{equation}
which has the formal solution
\begin{equation}
\label{e:evolution_operator}
u(\tau,\sigma)=e^{iL\tau}u(0,\sigma)
\end{equation}
in terms of the (non-unitary: see Sec.~\ref{s:energy_norm} below) evolution operator $e^{iL\tau}$. By further performing a harmonic decomposition $u(\tau,\sigma)\sim u(\sigma)e^{i\omega\tau}$ in Eq.~\eqref{matrix evolution} we arrive at the eigenvalue equation
\begin{equation}\label{eigenvalue problem}
L u_n=\omega_n u_n,
\end{equation}
where $\omega_n$ is an infinite set of eigenvalues of the operator $L$, with $n\geq 0$ the mode number. Thus, the calculation of QNMs through the hyperboloidal framework ultimately translates to the eigenvalue problem of the operator $L$, which, in turn, contains information concerning the boundary conditions and the spacetime metric. Finally, since $\partial_t=(1/\lambda)\partial_\tau$, $t$ and $\lambda\tau$ ``tick'' at the same rate. Therefore, the QNMs $\omega_n$ conjugate to the two distinct temporal coordinates coincide up to a scaling constant $1/\lambda$, and the change in the time coordinate does not affect the QNM frequencies~\cite{Jaramillo:2020tuu}.

\subsubsection{Hyperboloidal framework in the Reissner-Nordstr\"om spacetime}\label{hyperboloidal}

In what follows, we choose the BH horizon as the characteristic length scale,\footnote{Note that $\lambda = 2r_+$ in Ref.~\cite{PanossoMacedo:2018hab}. This implies that the quantities $\rho(\sigma)$ and $h(\sigma)$ differ from those in~Ref.~\cite{PanossoMacedo:2018hab} by a factor of $2$.} so that $\lambda = r_+$. Ref.~\cite{PanossoMacedo:2018hab} introduces the so-called minimal gauge, in which a compactification in the radial coordinate follows from
\bea
\label{radial compactification}
r=r_+ \dfrac{\rho(\sigma)}{\sigma}, \quad \rho(\sigma) = 1 -\rho_1(1-\sigma),
\eea
where $\rho_1$ is a parameter yet to be fixed. With this choice, the horizon $r_+$ is fixed at $\sigma=1$ regardless of $\rho_1$. As expected, $r\rightarrow \infty$ is mapped to $\sigma=0$. Substituting Eq.~\eqref{radial compactification} into the the tortoise coordinate \eqref{tortoise RN} leads to
\bea
\label{Tortoise RN 2}
\dfrac{r_*}{r_+} &=& \rho_1 + \dfrac{\ln(1-\rho_1)}{1-\kappa^2} + \dfrac{1-\rho_1}{\sigma} - (1+\kappa^2)\ln\sigma \nn \\
&+& \dfrac{ \ln(1-\sigma) - \kappa^4 \ln \left[ 1-\rho_1 - \sigma (\kappa^2 - \rho_1)\right]}{1-\kappa^2}.
\eea
The right-hand side of Eq.~\eqref{Tortoise RN 2} provides the overall structure for the function $g(\sigma)$. In particular, one can ignore any term not depending on $\sigma$ when defining $g(\sigma)$, because only $g'(\sigma)$ contributes to the expressions in Eq.~\eqref{new operators}. Finally, the height function in the minimal gauge reads
\bea
&h(\sigma) = g(\sigma) + h_0(\sigma), \nn \\
&h_0(\sigma) = 2\left[ (1+\kappa^2) \ln\sigma - \dfrac{1-\rho_1}{\sigma}\right].
\label{RN height}
\eea
There is still a remaining degree of freedom within the minimal gauge, that is, the choice of the parameter $\rho_1$. Below, we will describe the two available options providing us with different limits to extremality~\cite{PanossoMacedo:2018hab}: the usual extremal RN spacetime, and the near-horizon geometry given by the Robinson-Bertotti solution~\cite{Robinson59,Bertotti59}.

\smallskip
{\emph{\bf Areal radius fixing gauge.}} The simplest choice is to set $\rho_1=0$, so that $r=r_+/\sigma$. We refer to this case as the {\em areal radius fixing gauge}, since it implies $\rho(\sigma)=1$ in the above expressions. It then follows that the Cauchy horizon $r_-$ is located at $\sigma_- = \kappa^{-2}$, i.e. its location in the new compact coordinate $\sigma$ changes with the charge parameter $\kappa$. In particular, the Schwarzschild limit $\kappa=0$ yields the BH singularity $\sigma_-\rightarrow \infty$ ($r_- = 0$), whereas the event and Cauchy horizons coincide at $\sigma_+=\sigma_-=1$ in the extremal limit $|\kappa|=1$. 

By fixing $\rho_1=0$, Eqs.~\eqref{Tortoise RN 2} and \eqref{RN height} lead to
\begin{align}\label{hyperboloidal1}
g(\sigma) &= \dfrac{1}{\sigma} - (1+\kappa^2) \ln\sigma + \dfrac{\ln(1-\sigma) -\kappa^4\ln(1-\kappa^2\sigma)}{1-\kappa^2}, \\ \label{hyperboloidal2}
h(\sigma) &=  -\dfrac{1}{\sigma} + (1+\kappa^2) \ln\sigma + \dfrac{\ln(1-\sigma) -\kappa^4\ln(1-\kappa^2\sigma)}{1-\kappa^2}.
\end{align}
Inserting Eqs.~\eqref{hyperboloidal1} and \eqref{hyperboloidal2} into Eqs.~\eqref{new operators} yields
\begin{align}
p(\sigma) &= \sigma^2(1-\sigma)(1-\kappa^2\sigma), \\	
w(\sigma) &= 4 \left[ 1 + \kappa^2(1+\kappa^2)(1-\sigma)\right] \left[1 + \sigma (1+\kappa^2)\right], \\
\gamma(\sigma) &= 1 - 2 \left[ 1 + \kappa^2(1+\kappa^2) \right]\sigma^2 + 2\kappa^2(1+\kappa^2) \sigma^3,\\
q(\sigma) &= \ell(\ell+1) +\sigma \left( \mu - \kappa^2 \nu \sigma \right), 
\end{align}
which completely characterize the operator $L$ and the spectral problem of scalar and gravitoelectric perturbations in the RN spacetime. Here, the singular character of the operator $L$ becomes more evident, since $p(\sigma)$ clearly vanishes at $\sigma=0$ ($\scri^+$) and $\sigma=1$ (${\cal H}^+$), as well as at the Cauchy horizon $\sigma=\kappa^{-2}$.

From the differential equation perspective, it is important to note the different singular character of future null infinity and that of the event and Cauchy horizons. More specifically, $p(\sigma)$ behaves as $\sigma^2$ when $\sigma\rightarrow 0$, though it vanishes linearly as $\sigma \rightarrow 1$ or $\sigma \rightarrow \kappa^{-2}$, as long as $|\kappa|\neq 1$. In other words, in the subextremal case the operator $L$ possesses an essential singularity (irregular singular point) at future null infinity ($\sigma = 0$), whereas it has a removable singularity (regular singular point) at the horizons. 

In the extremal case $|\kappa|=1$, however, both horizons coincide. As a consequence, $p(\sigma)$ also possesses an essential singularity at $\sigma=1$, i.e., it vanishes as $(1-\sigma)^2$. This property is a direct manifestation of the so-called discrete conformal isometry between the extremal horizon and spacetime boundaries at infinity~\cite{Lubbe:2013yia},
namely the duality between spatial infinity $i^0$ and the horizon bifurcation sphere, on the one hand, and between null infinity and the regular part of the horizon, on the other hand. Such a symmetry explains, for instance, the Aretakis instability of a massless scalar field at the extremal RN horizon~\cite{Aretakis:2011ha,Aretakis:2011hc} in terms of well-known results for the field's decay at future null infinity~\cite{Bizon:2012we}.

To better identify this symmetry in the extremal limit $|\kappa|=1$, it is convenient to map the radial coordinate $\sigma \in [0,1]$ into $x\in[-1,1]$ via the transformation
\beq
x = 2\sigma -1, \quad \sigma = \dfrac{1+x}{2}.
\eeq
Hence, the conformal isometry between the extremal horizon and spacetime boundaries at infinity is assessed via the mapping $x\rightarrow - x$. In the limit $|\kappa| \rightarrow 1$, the height and compactification functions in Eqs.~\eqref{hyperboloidal1} and \eqref{hyperboloidal2} read, in terms of the coordinate $x$:
\begin{align}
g(x) &= \dfrac{2}{1+x} - \dfrac{1+x}{1-x} - 2 \ln(1+x) +  2 \ln(1-x), \\
h(x) &= -\dfrac{2}{1+x} + \dfrac{1+x}{1-x} - 2 \ln(1+x) +  2 \ln(1-x),
\end{align} 
which in turn leads to the following form for the functions entering the operator $L$ [cf.~Eq.~\eqref{new operators}]:
\beq
\label{extremal functions}
p(x)=\dfrac{(1-x^2)^2}{8},\,\, w(x)=2(4-x^2), \,\,
 \gamma(x) = -\dfrac{x(3-x^2)}{2}.
\eeq
The symmetry of $p$ and $w$ as $x\rightarrow -x$ is evident.\footnote{Note that $p$ is actually associated with a second-order operator $\sim \partial^2_x$, which is also symmetric under $x\rightarrow-x$.} The odd symmetry of $\gamma \rightarrow -\gamma$ is in accordance with its definition in terms of the operator $L_2$. Indeed, $L_2$ remains invariant since $\partial_x \rightarrow - \partial_x$. Clearly, the symmetry of the operator $L_1$ ultimately depends on the behavior of the potential $q(x)$. We will study this feature in Sec.~\ref{sec Gravitoelectric}.

\smallskip
{\emph{\bf Cauchy horizon fixing gauge.}} A second option allows us to fix the Cauchy horizon $\sigma_-$ at a coordinate location that does not depend on the parameter $\kappa$. In particular, the choice $\rho_1 = \kappa^2$ fixes the Cauchy horizon at $\sigma_- \rightarrow \infty$~\cite{PanossoMacedo:2018hab}. We call this the {\em Cauchy horizon fixing gauge}. In the extremal limit $|\kappa|\rightarrow 1$, this gauge shows a discontinuous transition in the near-horizon geometry~\cite{PanossoMacedo:2018hab}.

With the choice $\rho_1 = \kappa^2$, and ignoring terms not depending on $\sigma$, one reads from Eqs.~\eqref{Tortoise RN 2} and \eqref{RN height}
\bea
g(\sigma) &=&  \dfrac{1-\kappa^2}{\sigma} - (1+\kappa^2)\ln\sigma + \dfrac{ \ln(1-\sigma)}{1-\kappa^2}, \\
h(\sigma) &=& - \dfrac{1-\kappa^2}{\sigma} + (1+\kappa^2)\ln\sigma + \dfrac{ \ln(1-\sigma)}{1-\kappa^2}.
\eea
From these expressions, Eqs.~\eqref{new operators} yield
\bea
&& p(\sigma) = (1-\kappa^2)\dfrac{\sigma^2(1-\sigma)}{\rho(\sigma)^2},  \nn \\
&& q(\sigma) = \dfrac{1-\kappa^2}{\rho(\sigma)^2} \left[ \ell(\ell+1) + \dfrac{\sigma}{\rho(\sigma)}\left( \mu - \kappa^2 \nu \dfrac{\sigma}{\rho(\sigma)}\right)\right], \nn \\
&& w(\sigma) = 4\dfrac{\sigma + \rho(\sigma)}{\rho(\sigma)^2}, \quad \gamma(\sigma) =1-\dfrac{2\sigma^2}{\rho(\sigma)^2}. \nn \\
\eea
The singular behavior at the extremal limit $|\kappa|=1$ is evident in the functions $p(\sigma)$ and $q(\sigma)$, since $L_1$ vanishes altogether. A regularization can however be implemented by rescaling $\tau$ with a $(1 - \kappa^2)$ factor: this is relevant in the Robinson-Bertotti solution limit (cf.~\cite{PanossoMacedo:2018hab}).

\section{Pseudospectrum}\label{pseudospectra}

Following Ref.~\cite{Jaramillo:2020tuu}, we start by discussing the intuitive notion of spectral stability, in which a perturbation of order $\epsilon$ to the underlying operator leads to perturbed QNMs migrating up to a distance of the same order $\epsilon$. This result is formally proven in the context of self-adjoint operators. More specifically, if one considers a linear operator $L$ on a Hilbert space $\mathcal{H}$ with a scalar product $\langle \cdot , \cdot \rangle$, then the adjoint $L^\dagger$ is the linear operator fulfilling $\langle L^\dagger u,v\rangle=\langle u,Lv\rangle$, for all $u, v$ in $\mathcal{H}$. The operator $L$ is normal if and only if $\left[L,L^\dagger\right]=0$. Clearly, self-adjoint operators $L=L^{\dagger}$ are normal. In this context, the spectral theorem for normal operators ensures that eigenfunctions form an orthonormal basis and that the eigenvalues are stable under perturbations of $L$. 

On the contrary, non-normal operators lack a spectral theorem and lead to a weak control of eigenfunction completeness and eigenvalue stability. Thus, the analysis based solely on the spectrum may be misleading if the system is beset with spectral instability, where the perturbed eigenvalues may extend into far regions in the complex plane, despite a rather small change in the operator. Indeed, strong non-normality, occurring in quite generic settings, leads to a severely uncontrolled spectrum under perturbations of the governing operator~\cite{Trefethen:1993,Driscoll:1996,Davies99,Trefethen:2005,Davie07,Sjostrand2019,Jaramillo:2020tuu}. 

The notion of pseudospectrum formally captures the sensitivity of the spectrum to perturbations~\cite{Trefethen:2005,Davie07,Sjostrand2019}. To introduce this concept, we first recall the definition of an eigenvalue as the values $\omega$ for which $\omega \mathbb{I}-L$ is singular. To intuitively enlarge this notion, one may inquire: what is the region in the complex plane in which $||\omega \mathbb{I}-L||$ is small, or equivalently, in which $||\omega \mathbb{I}-L||^{-1}$ is large?
 
Addressing this question naturally leads to the definition of the pseudospectrum~\cite{Trefethen:2005,Davie07,Sjostrand2019}, which states that the $\epsilon$-pseudospectrum $\sigma^\epsilon(L)$ with $\epsilon> 0$ is the set of $\omega\in \mathbb{C}$ for which $||\left(\omega\mathbb{I}-L\right)^{-1}||>\epsilon^{-1}$. The operator $\left(\omega\mathbb{I}-L\right)^{-1}$ is called the resolvent of $L$ at $\omega$. In turn, $||\left(\omega\mathbb{I}-L\right)^{-1}||$ diverges for $\omega\in\sigma(L)$, where $\sigma(L)$ is the spectrum of $L$, so that (from the very definition of the pseudospectrum) the spectrum -- a discrete set of numbers in the complex plane -- is contained in the $\epsilon$-pseudospectrum for every $\epsilon$. In other words, the $\epsilon$-pseudospectra  $\sigma^\epsilon(L)$ are nested sets around the spectrum. Specifically, the norm of the resolvent maps values $\omega$ in the complex plane into real positive numbers, in such a way that the boundaries of $\sigma^\epsilon(L)$ are contour levels of the function defined by the norm of the resolvent. Such boundaries are then nested contour lines around the spectral points. As $\epsilon\rightarrow0$, $\sigma^\epsilon(L)\rightarrow\sigma(L)$. The structure of the resolvent encodes fundamental information concerning non-trivial structures in the complex plane that cannot be revealed by the spectrum alone. For instance, normal operators exhibit trivial resolvent structures that extend circularly up to order $\sim\epsilon$ around the spectrum (e.g.~Fig.~4 in Ref.~\cite{Jaramillo:2020tuu}). Non-normal operators may possess resolvent structures that extend far from the spectrum, which is an imprint of poor analytic behavior of the resolvent as a function of $\omega$.

The connection between the low regularity/analyticity of the resolvent and the underlying spectral instabilities follows from considering a perturbed operator $L+\delta L$, with perturbation norm $||\delta L||<\epsilon$. If the spectrum of the perturbed operator stays bounded in a vicinity of order $\sim\epsilon$ around $\sigma(L)$, then the operator displays spectral stability. On the other hand, if it migrates in the complex plane far from $\sigma(L)$ at distances that are orders of magnitude larger than $\epsilon$, then $L$ is spectrally unstable. This distinction can be made directly through the pseudospectrum, at the level of the non-perturbed operator $L$, without the need of systematically introducing perturbations to the operator.

We emphasize that the definition of the pseudospectrum, and therefore any statement on spectral instability, depends on the choice of the underlying scalar product $\langle \cdot , \cdot \rangle$. The scalar product fixes the norm that quantifies the notion of ``big'' or ``small'' perturbations $||\delta L||$. On physical grounds, and following~\cite{Driscoll:1996,Jaramillo:2020tuu}, we argue in favor of the system's energy as the most adequate norm to control the pseudospectra and assess spectral instability, because it conveniently encodes the size of the physical perturbation with respect to the nature of the problem (see \cite{Gasperin:2021kfv} for a systematic discussion of this point). On the computational side, calculating the pseudospectrum requires the numerical evaluation of the norm of the resolvent, according to the previous definition. The numerical scheme must consistently incorporate the particular choice for the norm. The next section summarizes the core concepts needed in this work.

\subsection{The energy norm}
\label{s:energy_norm}
A physically motivated choice is the so-called energy norm~\cite{Driscoll:1996,Jaramillo:2020tuu}, a natural way of framing the problem in terms of the physical energy contained in the field $\phi$ with dynamics dictated by the wave equation~\eqref{evolution equation}. Within the hyperboloidal formulation of the wave equation given by Eq.~\eqref{matrix evolution}, the energy norm for a ($\ell$-mode) vector $u$ reads~\cite{Jaramillo:2020tuu} (see \cite{Gasperin:2021kfv} for a full account of its relation with the total energy of the field on a spacetime slice)
\begin{align}\label{energy norm}
&||u||^2_{_{E}} = \Big|\Big|\begin{pmatrix}
	\phi \\
	\psi
\end{pmatrix}\Big|\Big|^2_{_{E}} \equiv E(\phi, \psi) \\
&=\frac{1}{2}\int_0^1 \left(w(\sigma)|\psi|^2
+ p(\sigma)|\partial_\sigma\phi|^2 + q(\sigma) |\phi|^2\right) d\sigma, \nn
\end{align}
where the subscript $E$ denotes the energy norm, and the integration limits correspond to the $[0,1]$ compact spatial interval in our hyperboloidal scheme. In turn, the energy scalar product that defines the norm reads
\bea
&&	\langle u_1,\! u_2\rangle_{_{E}} = \Big\langle\begin{pmatrix}
	\phi_1 \\
	\psi_1
\end{pmatrix}, \begin{pmatrix}
	\phi_2 \\
	\psi_2
\end{pmatrix}\Big\rangle_{_{E}}  \\
&&=
\frac{1}{2} \int_0^1 \left( w(\sigma)\bar{\psi}_1 \psi_2 + p(\sigma)  \partial_\sigma\bar{\phi}_1\partial_\sigma\phi_2 + q(\sigma)\bar{\phi}_1 \phi_2 \right)  d\sigma, \nn\label{energy scalar product}
\eea
from which $||u||^2_{_{E}} = \langle u, u\rangle_{_{E}}$ trivially holds. Due to the dissipative nature of the block operator $L_2$ --- see Eqs.~\eqref{L2} and \eqref{matrix evolution} --- which encodes the boundary conditions at null infinity and at the event horizon, the operator $L$ is not selfadjoint with respect to the scalar product \eqref{energy scalar product}~\cite{Jaramillo:2020tuu}: this justifies the non-unitary character of the evolution operator in Eq.~(\ref{e:evolution_operator}). We use Chebyshev spectral methods to numerically implement Eq.~\eqref{energy scalar product} in the calculation of the pseudospectrum.

\subsection{Chebyshev's spectral method}

There are various methods to compute BH QNMs~\cite{Leaver85,Iyer:1986np,Berti:2009kk,Konoplya:2011qq}. The differential operator \eqref{eigenvalue problem} can be discretized and turned into a matrix problem using Chebyshev's spectral method~\cite{Trefethen:2000,Trefethen:2005}.

The compactified domain of the operator $L$, $\sigma\in\left[0,1\right]$, is discretized with $N+1$ Chebyshev-Lobatto interpolation grid points, while for the discretization of the differential operators we utilize Chebyshev differentiation matrices~\cite{Trefethen:2000}. Overall, the resulting matrix $L$ has dimensions $\mathcal{N}\times \mathcal{N}$, with $\mathcal{N}=2(N+1)$, where the factor of $2$ comes from the first-order reduction in time, so $(N+1)$ values are used for each $\phi$ and $\psi$. Once $L$ is discretized, the BH QNMs follow straightforwardly from the eigenvalues of the matrix.

To calculate matrix norms and pseudospectra, one must also implement the Chebyshev-discretized version of the energy scalar product norm~\eqref{energy scalar product} via
\begin{equation}
\langle u, v\rangle_{_E} = (u^*)^i G^{E}_{ij}v^j = u^*\cdot G^{E}\cdot v, \ \ u,v \in \mathbb{C}^\mathcal{N},
\end{equation}
with $u^*$ the conjugate transpose of $u$. The construction of the Gram matrix $G^E_{ij}$ corresponding to \eqref{energy scalar product} is detailed in Appendix A of Ref.~\cite{Jaramillo:2020tuu}, and the adjoint operator $L^\dagger$ reads
\begin{equation}\label{adjoint}
L^\dagger = \left(G^E\right)^{-1}\cdot L^*\cdot G^E.
\end{equation}
Finally, the $\epsilon$-pseudospectrum $\sigma^\epsilon_E(L)$ in the energy norm is given by~\cite{Jaramillo:2020tuu}
\begin{equation}
\label{pseudospectra energy norm}
\sigma^\epsilon_{_E} (L) = \{\omega\in\mathbb{C}: s_{_E}^\mathrm{min}(\omega \mathbb{I}- L)<\epsilon\}, 
\end{equation}
where $s_E^\mathrm{min}$ is the minimum of the generalized singular value decomposition, which incorporates the adjoint in the energy scalar product
\begin{equation}
\label{svd definition}
s_{_E}^\mathrm{min}(M) = \min \{\sqrt{\omega}:  \omega\in \sigma(M^\dagger M) \}, \quad M=\omega \mathbb{I}- L.
\end{equation}

\subsection{QNM-free regions: logarithmic boundaries and pseudospectra}\label{QNM free region}
We conclude our general discussion of the pseudospectrum with a short summary of ``universality classes''~\cite{trefethen2005spectra,zworski2017mathematical,Sjostrand2019}. As discussed in Sec.~\ref{pseudospectra}, the $\epsilon$-pseudospectrum $\sigma^\epsilon(L)$ determines the maximal region in the complex plane that QNMs can reach under perturbations of norm $\epsilon$ on the operator $L$. Equivalently, the regions beyond a given $\epsilon$-pseudospectrum boundary are called QNM-free (or resonance-free) regions. 

The study of the asymptotics of QNM-free regions is crucial to assess the existence of QNM-free strips around the real axis, so that the notion of fundamental (or principal) resonance, understood as the closest to the real axis, makes sense. This is fundamentally related to the study of the decay of waves scattered by a resonator (in our case, the BH), and therefore is key in the context of GW ringdown.

The QNM-free regions fall into different ``universality classes'' according to their asymptotic behavior for large real parts of the frequencies:
\bea
\label{e:resonance_free_regions_1}
\mathrm{Im}(\omega) < F(\mathrm{Re}(\omega)), \ \ \mathrm{Re}(\omega)\gg 1,
\eea
where $F(x)$ is a real function controlling the asymptotics. The mathematical literature (cf. e.g. \cite{zworski2017mathematical,dyatlov2019mathematical}) identifies several possibilities, such as
\bea
\label{e:resonance_free_regions_2}
F(x) =
\left\{
\begin{array}{rcl}
&\hbox{(i)}& \ e^{\alpha x}, \ \alpha>0 \\
&\hbox{(ii)}& \ \tilde{C}, \\
&\hbox{(iii)}& \ C\ln(x), \\
&\hbox{(iv)}& \ \gamma \; x^\beta,\, \beta\in\mathbb{R},\, \gamma> 0
\end{array}
\right. 
\eea
where $C, \tilde{C}, \alpha, \beta, \gamma$ are constants. The specific form of $F$ and the constants in Eqs.~(\ref{e:resonance_free_regions_2}) depend on qualitative features of the underlying effective potential $V$ and on the boundary conditions of the problem. Typical behaviors in our setting belong to cases (iii) or (iv) (see~\cite{zworski2017mathematical}). In particular, logarithmic resonant-free regions of class (iii) appear in the setting of generic scattering by impenetrable obstacles or by potentials (either of compact support or extending to infinity) when we allow for low regularity~\cite{Regge58,LaxPhi71,LaxPhi89,Vainb73,zworski2017mathematical,dyatlov2019mathematical,Sjoes90,Marti02,SjoZwo07}. In the case of potentials and/or boundary conditions with enhanced regularity, the more stringent power-law (iv) controls the QNM-free regions.

For the Schwarzschild potential, numerical investigations~\cite{Jaramillo:2021tmt} have shown that the QNM-free regions are asymptotically bounded from below by logarithmic curves of the form
\bea
\label{e:log_branches}
\mathrm{Im}(\omega) \sim C_1 + C_2 \ln \big[\mathrm{Re}(\omega) + C_3\big].
\eea
In the asymptotic regime $\mathrm{Re}(\omega)\gg 1$ the constants $C_1$ and $C_3$
can be neglected, leading to case (iii) in Eq.~(\ref{e:resonance_free_regions_2}). Nevertheless, it is remarkable that by adding nonzero constants $C_1$ and $C_3$ the logarithmic behavior holds also at intermediate values of $\mathrm{Re}(\omega)$ and even close to the actual QNM spectrum (in a region one can hardly consider as asymptotic).
These features will be investigated in detail below, confirming the preliminary results in~\cite{Jaramillo:2021tmt}
and, more importantly, providing a systematic extension and refinement that is crucial for the assessment of asymptotic universality.

\section{The pseudospectrum of Reissner-Nordstr\"om black holes}

In what follows we discuss the pseudospectrum of subextremal and extremal RN BHs, calculated (unless specified otherwise) within the areal radius fixing gauge. To ensure agreement with the known values of the QNM frequencies~\cite{Chandrasekhar:1985kt,Kokkotas:1988fm,Richartz:2014jla,Richartz:2015saa} we typically use $N=200$ grid points and set the internal precision in all calculations to $20\times\text{MachinePrecision}\sim 300$ digits. We scanned the part of the complex plane shown in our plots with resolution $\sim250\times 150$, corresponding to $\sim 3.8\times 10^4$ points.

\begin{figure}[t]\hspace{-0.51cm}
	\includegraphics[scale=0.34]{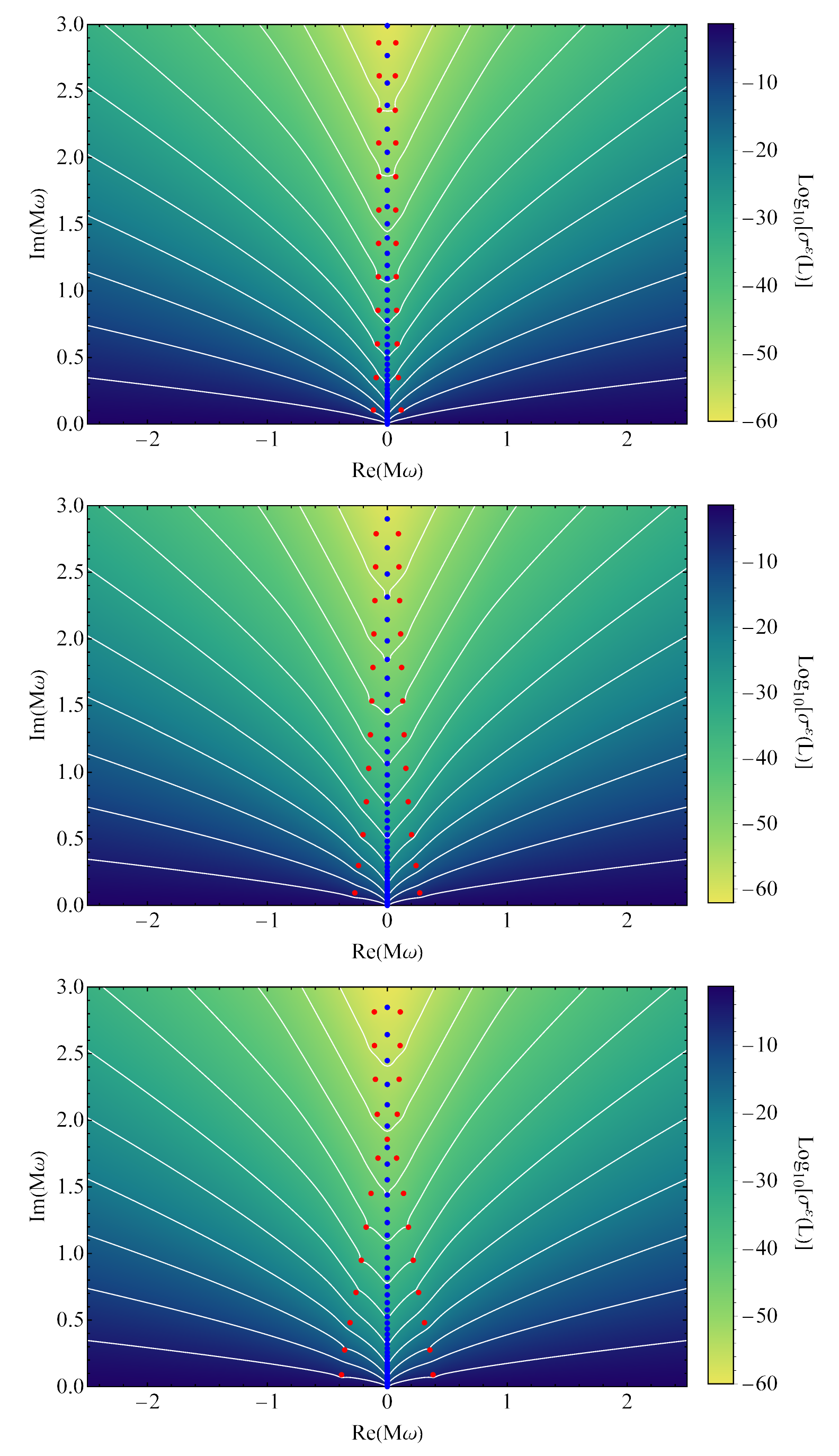}	
	\caption{Pseudospectra of a BH with charge $Q/M=0.5$. Top: $\ell=0$ scalar QNMs (red dots) and $\epsilon$-pseudospectra boundaries (white lines). Middle: $\ell=1$ electromagnetic-led QNMs and $\epsilon$-pseudospectra boundaries. Bottom: same for $\ell=2$ gravitational-led QNMs. The contour levels $\log_{10}(\epsilon)$ range from $-55$ (top level) to $-5$ (bottom level) in steps of $5$. The blue dots designate branch-cut (non-convergent) modes.}
	\label{grel}
\end{figure}

In the asymptotically flat setting of RN spacetime and due to the inclusion of (non-regular) null infinity in the grid of the compactified hyperboloidal formulation, the counterpart of the power-law decay of asymptotically flat effective potentials -- which is responsible for the branch cut in the Green's function of Eq.~\eqref{evolution equation}~\cite{Leaver:1986gd} and for the polynomial late-time tails~\cite{Price:1971fb,Gundlach:1993tp,Gundlach:1993tn} -- is seeded into the operator $L$. This means that the operator $L$ contains both the discrete set of QNMs (the ``point spectrum'') and a continuous spectrum. 

With the discretized approximate operator $L$, the continuous spectrum turns into a discrete set of points along the positive imaginary axis. In contrast to QNMs, which converge to known values~\cite{Chandrasekhar:1985kt,Kokkotas:1988fm,Richartz:2014jla,Richartz:2015saa}, these additional ``discrete'' eigenvalues of the approximate operator $L$ do not converge as $N\rightarrow\infty$. On the contrary, they accumulate on the positive imaginary axis along the expected branch cut. We will refer to these points as ``branch-cut'' modes. We must take these points into account when calculating pseudospectra, since they are fundamentally entangled with the discretized formulation of the spectral problem. For clarity we visualize them as blue points in the figures that follow (in contrast to converging QNMs, shown as red points).

\subsection{Spectral instability: the pseudospectra}

In this section we discuss, for illustration, the pseudospectrum of BHs with charge $Q/M=0.5$. Our results for scalar $\ell=0$, electromagnetic-led $\ell=1$ and gravitational-led $\ell=2$ perturbations are shown in Fig.~\ref{grel}  (note the different units for the frequency, as compared with the Schwarzschild case in \cite{Jaramillo:2020tuu}, where a normalization $4M\omega$ is used). More specifically, Fig.~\ref{grel} displays both the spectra (red dots) and the underlying pseudospectra of the respective operators. As expected, the QNMs appear on the upper half of the complex plane, reflecting modal stability, and they concentrate in the vicinity of the imaginary axis. The pseudospectral contour lines, shown in white, correspond to different values of $\epsilon$, as shown in the log-scale color bar. 

Although the pseudospectra of non-normal operators, such as that shown in Fig.~\ref{grel}, still form circular sets if we zoom arbitrarily close to the spectrum, their large scale global structure presents open sets. These extend into large regions of the complex plane even for small $\epsilon$, indicating spectral instability. This property implies that small-scale perturbations $||\delta L||_E<\epsilon$ can lead to perturbed spectra which migrate into regions that are much further away than $\epsilon$. Such picture is in sharp contrast with Fig.~\ref{spectral_stability} of Appendix \ref{stability} (see also Fig. 4 in \cite{Jaramillo:2020tuu} and the discussion in \cite{Gasperin:2021kfv}), where nested circular sets of radius $\sim\epsilon$ form around non-perturbed QNMs. In that sense, we can differentiate between spectral stability and instability according to the topographic structure of pseudospectra. 

\begin{figure}
	\includegraphics[scale=0.39]{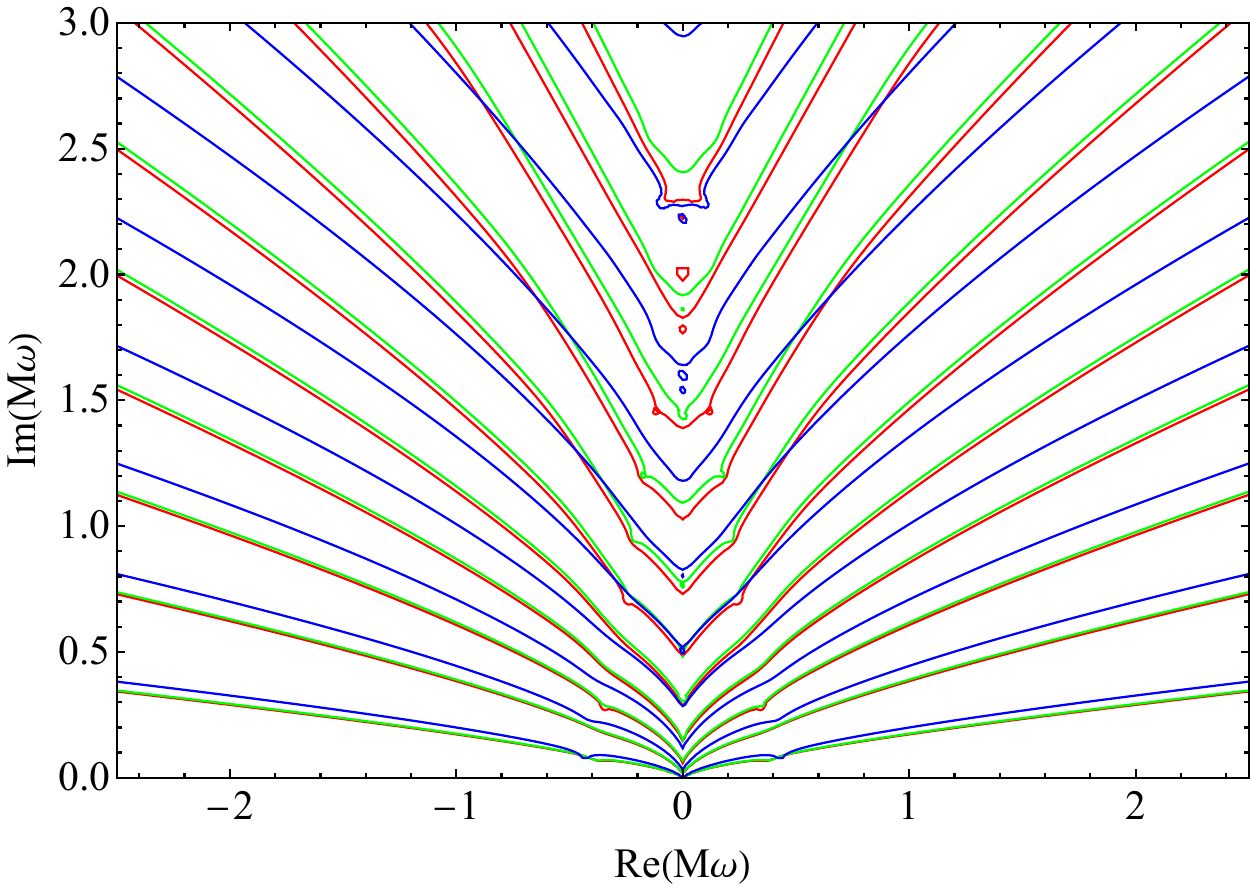}
	\caption{Comparison of pseudospectral levels for gravitational-led $\ell=2$ QNMs with varying BH charge. Here, $Q/M=0$, $0.5$ and $1$ correspond to red, green and blue contours, respectively. The red and green contour levels $\log_{10}(\epsilon)$ range from $-55$ (top level) to $-5$ (bottom level) in steps of $5$, while the blue ones from $-50$ (top level) to $-5$ (bottom level) in steps of $5$.}
	\label{Q_dependence}
\end{figure}

For the particular case of subextremal RN BHs, the structure of pseudospectra close to the imaginary axis depends on $\ell$. Higher angular indices move QNMs further away from the imaginary axis (higher real parts), as expected from the correspondence between QNMs and photon orbits~\cite{Berti:2005eb,Cardoso:2008bp}, and change the structure of the pseudospectra in a similar way. Asymptotically (beyond the region of non-perturbed QNMs), the pseudospectral boundaries have a logarithmic dependence, similar to that found in the Schwarzschild spacetime~\cite{Jaramillo:2021tmt}. We will return to this point in Sec.~\ref{asymptotics} below.

We observe a close similarity between the $\ell=2$ gravitational-led QNMs and pseudospectra in the bottom panel of Fig.~\ref{grel} and the $\ell=2$ gravitational QNMs and pseudospectra of the Schwarzschild spacetime (see Fig.~11 in~\cite{Jaramillo:2020tuu}, where $Q/M=0$). In Fig.~\ref{Q_dependence} we present a direct comparison of pseudospectral levels for selected values of $Q/M$. The pseudospectral contours for different BH charges are remarkably similar, consistently with the asymptotic universality discussed above, with a charge-dependent offset which increases as $Q/M\rightarrow1$ and as $\log_{10}(\epsilon)\rightarrow-\infty$.
\

\subsection{The areal radius and Cauchy fixing gauges}\label{sec:RadialCauchy}

Before discussing extremal geometries, it is important to address possible technical issues arising when taking the extremal limit. As discussed in Sec.~\ref{hyperboloidal}, the hyperboloidal areal radius fixing gauge and the Cauchy horizon fixing gauge have different limits to extremality, and the case $Q/M=1$ can only be studied within the areal radius fixing gauge.
If the pseudospectrum has a geometrical origin, it should depend only mildly on this gauge choice. We will show below that this is indeed the case by computing pseudospectra for subextremal BHs in two different gauges. Note that a similar issue appears when calculating the QNM spectrum in the extremal limit. For example, Leaver's continued fraction algorithm~\cite{Leaver90} is suitable for sub-extremal BHs, but it fails in a continuous limit to extremality. A modified version of Leaver's algorithm is only applicable if the extremal condition $Q/M=1$ is imposed at the onset of the calculation~\cite{Onozawa:1995vu,Richartz:2015saa}. 

These algorithms consider the problem in the frequency domain, and the strategy to incorporate the boundary conditions and regularize the underlying ordinary differential equation can be understood from a spacetime perspective in terms of the hyperboloidal  framework~\cite{PanossoMacedo:2018hab}. In particular, Ref.~\cite{PanossoMacedo:2018hab} shows that the Cauchy horizon fixing gauge naturally yields the regularization factor in the frequency domain employed by Leaver~\cite{Leaver90}. The technical failure in the algorithm to reach the extremal limit is then geometrically understood as a discontinuous transition to the near-horizon geometry. Similarly, the regularization factor in the frequency domain employed by Onozawa~\cite{Onozawa:1995vu} follows naturally if the wave equation is initially written in the areal radius fixing gauge, which has a well-behaved extremal limit.

\begin{figure}[t!]\hspace{-0.8cm}
	\includegraphics[scale=0.39]{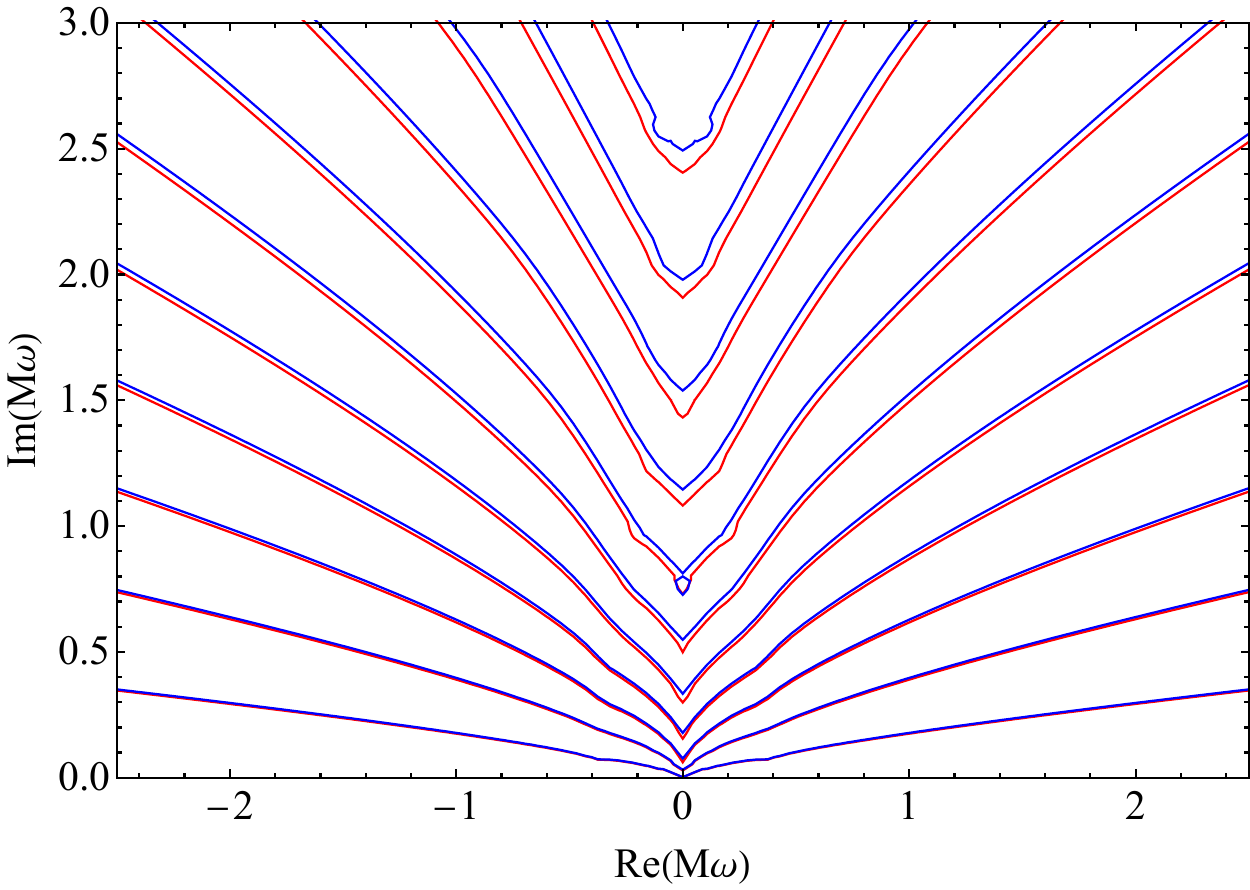}
	\caption{Comparison of the $\ell=2$ gravitational-led $\epsilon$-pseudospectra for a RN BH with $Q/M=0.5$. The pseudospectra in red were computed in the areal radius fixing gauge, while those in blue where computed in the Cauchy horizon fixing gauge. In both cases, the contour levels $\log_{10}(\epsilon)$ range from $-55$ (top level) to $-5$ (bottom level) in increments of $5$.}
	\label{radial_vs_Cauchy}
\end{figure}

\begin{figure}
	\includegraphics[scale=0.35]{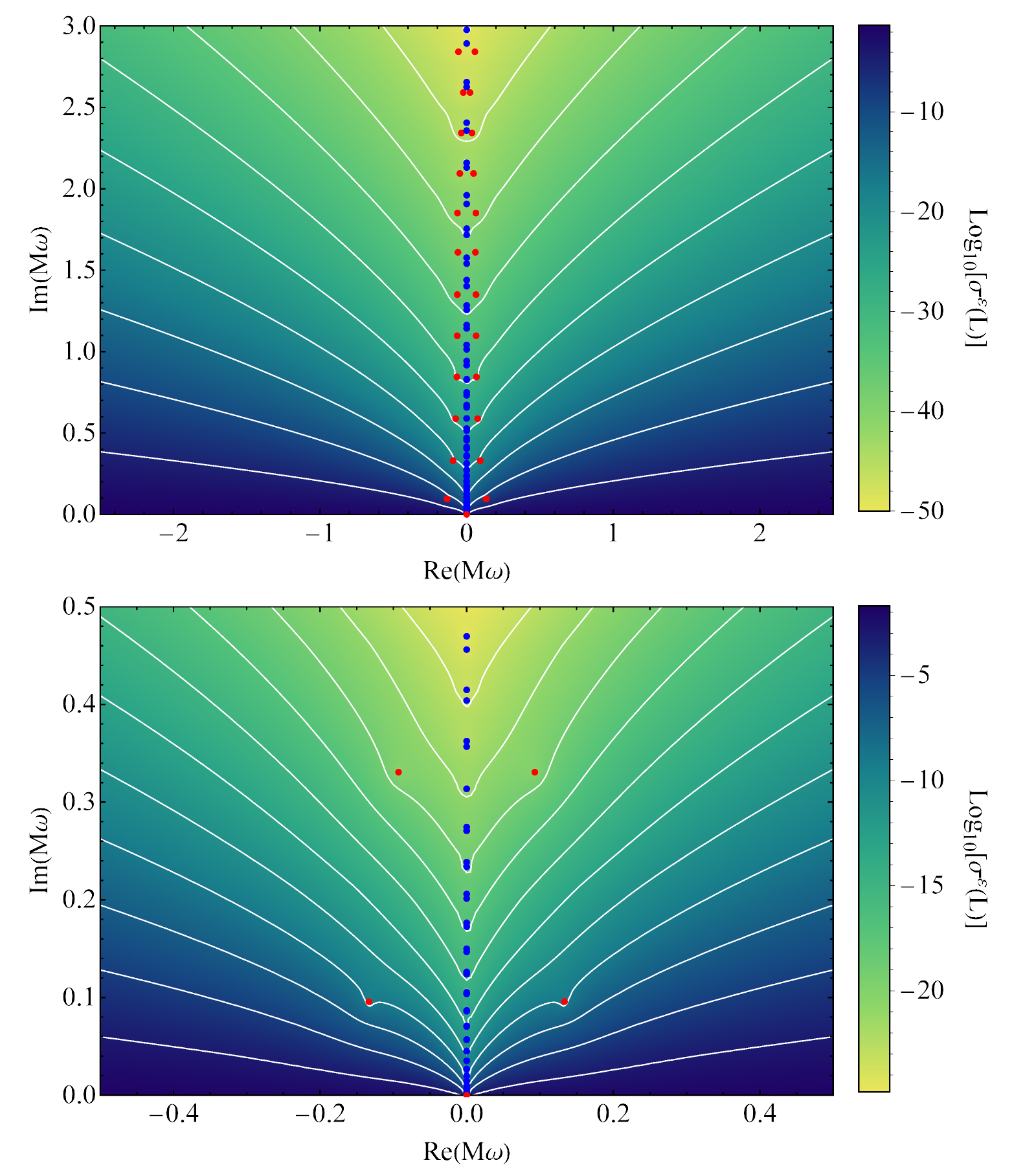}
	\caption{Top: $\ell=0$ scalar QNMs (red dots) and $\epsilon$-pseudospectra boundaries (white lines) of a RN BH with $Q/M=1$. The contour levels $\log_{10}(\epsilon)$ range from $-45$ (top level) to $-5$ (bottom level) in steps of $5$. Bottom: Enlarged region around the first few QNMs of the top panel. The contour levels $\log_{10}(\epsilon)$ range from $-23$ (top level) to $-3$ (bottom level) in steps of $2$. The blue dots designate branch-cut (non-convergent) modes.}
	\label{scalar}
\end{figure}

\begin{figure*}
	\includegraphics[scale=0.34]{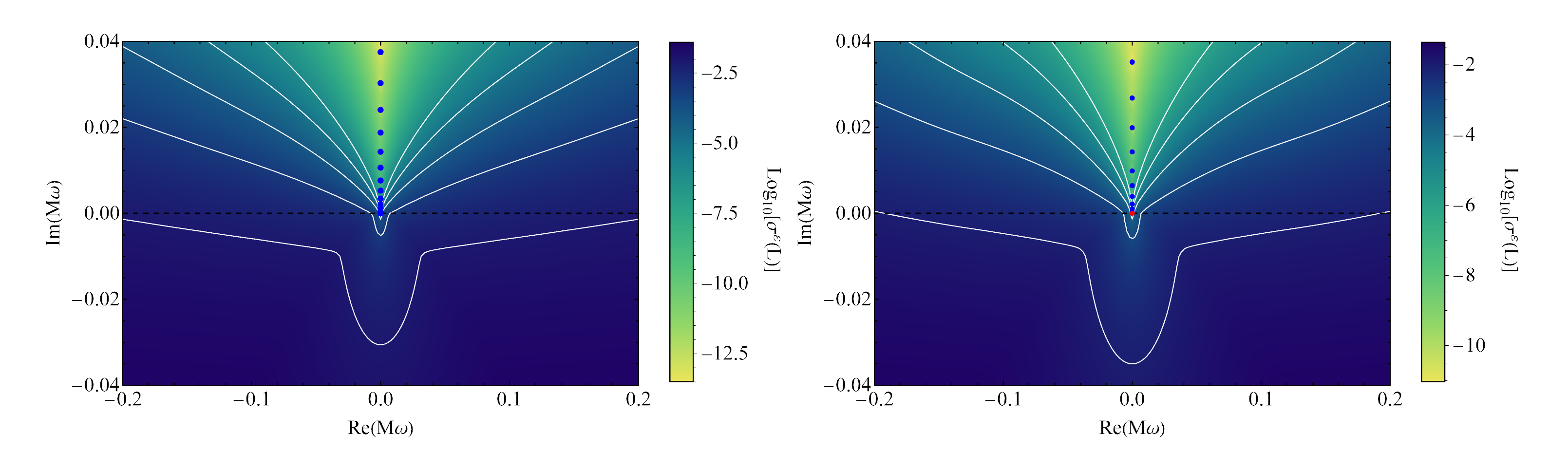}
	\caption{Left: enlargement of the region around the real axis (shown as a dashed black line) of $\ell=0$ scalar $\epsilon$-pseudospectral boundaries (white lines) for a RN BH with $Q/M=0.5$. Right: same, but for $Q/M=1$. A single zero-frequency QNM (shown as a red dot) exists in this case. The contour levels $\log_{10}(\epsilon)$ range from $-6$ (top level) to $-2$ (bottom level) in steps of $1$. The blue dots designate branch-cut (non-convergent) modes.}
	\label{scalar_zoom}
\end{figure*}

\begin{figure}\hskip -2ex
	\includegraphics[scale=0.39]{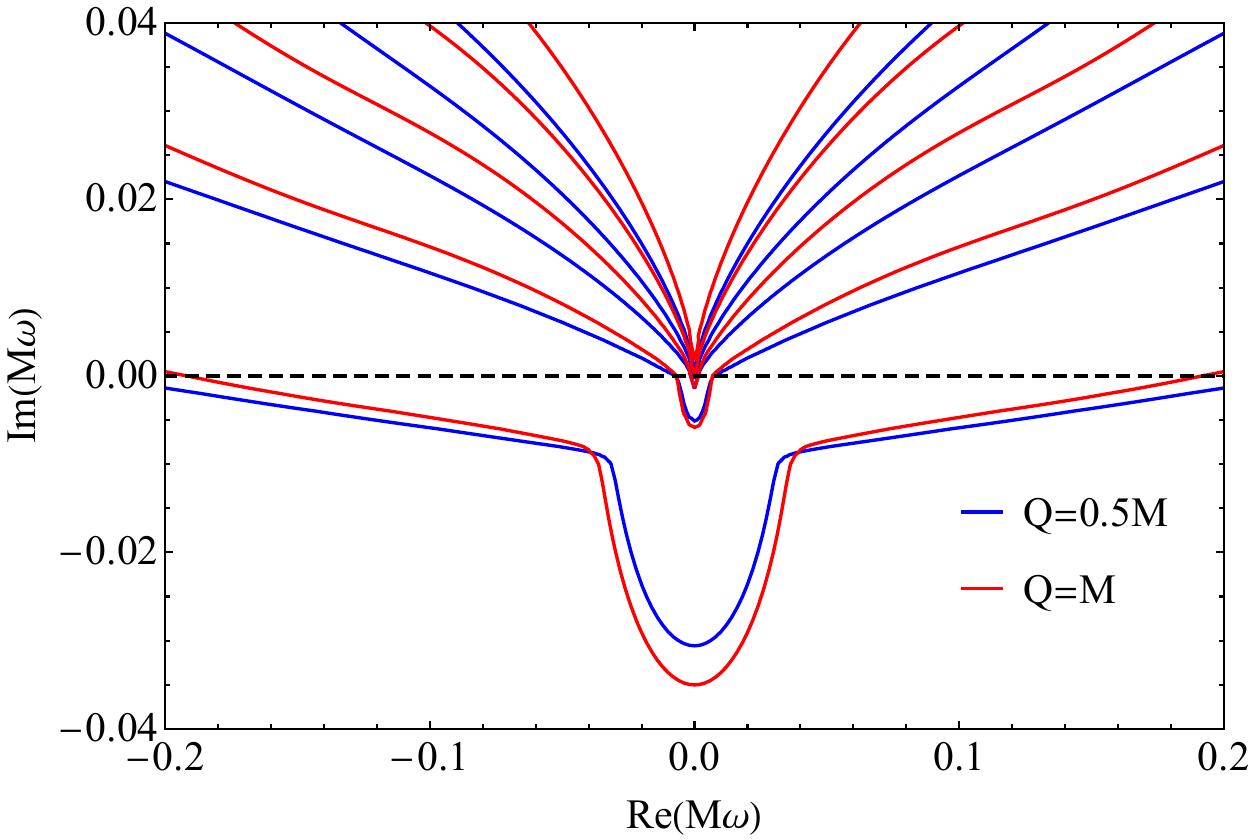}
	\caption{Overplotted contours from Fig.~\ref{scalar_zoom}, with the same parameters and contour levels.}
	\label{contour_overplot_zoom}
\end{figure}

Let us return to the issue of the slicing dependence of the pseudospectrum of subextremal RN BHs. Figure~\ref{radial_vs_Cauchy} shows the $\epsilon$-pseudospectra of gravitational-led QNMs for the two available hyperboloidal slices in a RN spacetime with $Q/M=0.5$, using exactly the same $\epsilon$ contour levels for both cases. We omit QNM frequencies from Fig.~\ref{radial_vs_Cauchy} for clarity. The pseudospectral contour lines reaching low overtones (large $\epsilon$) nearly coincide, but they separate as $\epsilon$ decreases. This can be understood in terms of a slight ``renormalization'' between the induced matrix norm in the areal radius and Cauchy fixing gauges, which is a consequence of using different functions in the construction of the operators $L_1$ and $L_2$. The (set of) pseudospectral contours are, to a very good approximation, gauge-independent, especially for small $\epsilon$. They have not only the same asymptotic logarithmic behavior (\ref{e:log_branches}) for both gauge choices, as expected, but also very similar values for the constants $C_1, C_2,$ and $C_3$. This demonstrates that the bulk properties of the pseudospectral levels do not depend on the choice of gauge, supporting the notion that the pseudospectrum is a geometrical property of the spacetime. 

\subsection{The extremal limit}\label{sec extremal}
\subsubsection{Scalar potential}

We first consider scalar fields in the extremal limit. The pseudospectrum is depicted in Fig.~\ref{scalar}, where we extend similar results presented in Fig.~\ref{grel}. The qualitative features of the pseudospectrum do not depend on the spin of the perturbing field. Note that the spectrum of extremal and near-extremal BHs is markedly different. In particular, near-extremal RN BHs have a family of slowly damped modes~\cite{Kim:2012mh,Zimmerman:2015trm,Richartz:2015saa,Cardoso:2017soq}. If we define $\delta \ll 1$ as
\beq
Q=M\left (1-\frac{\delta^2}{2}\right),
\eeq
then this family is well described by the purely damped mode 
\beq
M\omega=i \delta (n+\ell +1),\qquad n=0,\,1,\,2...
\eeq
Our numerical results agree with these predictions. For example, for $Q=0.999M$ $(\delta=0.0447214)$ we find the fundamental modes $M\omega=0.045125 i$ ($\ell=0$) and $M\omega=0.0899565 i$ ($\ell=1$). 

The pseudospectrum in the extremal case (see Fig.~\ref{scalar}), however, has the same qualitative log-like asymptotic behavior as in subextremal BHs, despite the existence of a zero-frequency QNM with $M\omega=0$, associated to the Aretakis instability~\cite{Aretakis:2011ha,Aretakis:2011hc,Angelopoulos:2018yvt}. The existence of a mode which lies at the origin of the complex plane is of key importance to transient instabilities that are typically resolved with pseudospectra: see e.g. the discussion of the transition to turbulence in hydrodynamics in Ref.~\cite{Trefethen:1993}. In the right panel of Fig.~\ref{scalar_zoom} we observe that the pseudospectral contour levels cross to the lower half of the complex plane for $\log_{10}(\epsilon)=-4$, or $\epsilon=10^{-4}$, which translates to an unstable perturbed spectrum. In fact, the transition occurs for similar levels in Ref.~\cite{Trefethen:1993}. This could be tantalizing evidence of analogies between BH spacetime instabilities and hydrodynamics. Unfortunately the evidence is inconclusive, because we observe a similar behavior for subextremal BHs (left panel of Fig.~\ref{scalar_zoom}). Contour lines of extremal RN dive slightly deeper into the unstable QNM region, as shown in Fig.~\ref{contour_overplot_zoom}, but we cannot draw definite conclusions due to the existence of the branch-cut eigenvalues, which ``poison'' the spectrum of the discrete operators and accumulate arbitrarily close to the origin. The present analysis is -- to our knowledge -- the first attempt to resolve transient BH instabilities through pseudospectra, but further work is required to draw solid conclusions.

\begin{figure*}
	\includegraphics[scale=0.36]{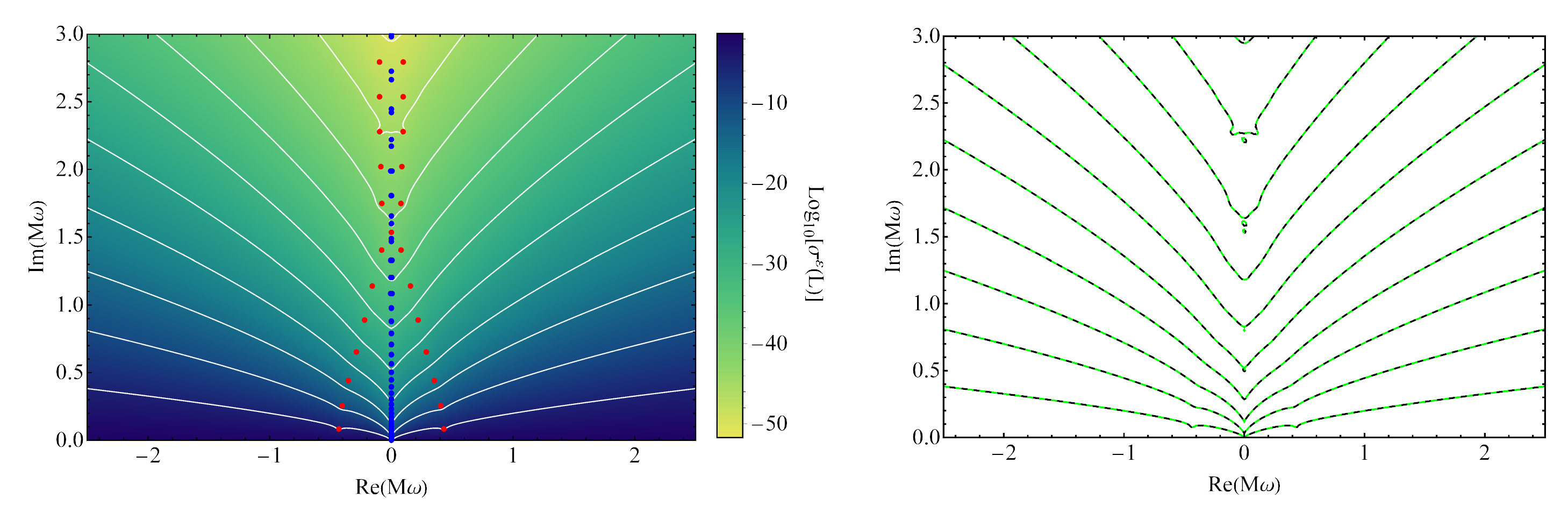}
	\caption{Left: $\ell=2$ gravitational-led QNMs (red dots) and $\epsilon$-pseudospectra boundaries (white lines) of a RN BH with $Q/M=1$. Right: superimposed $\ell=1$ electromagnetic-led (black lines) and $\ell=2$ gravitational-led (green dashed lines) $\epsilon$-pseudospectral contours of a RN BH with $Q/M=1$. In both cases, the contour levels $\log_{10}(\epsilon)$ range from $-50$ (top level) to $-5$ (bottom level) in steps of $5$. The blue dots designate branch-cut (non-convergent) modes.}
	\label{grelext}
\end{figure*}

\subsubsection{Gravitoelectric potential}\label{sec Gravitoelectric}

\begin{figure*}[t]
	\includegraphics[scale=0.42]{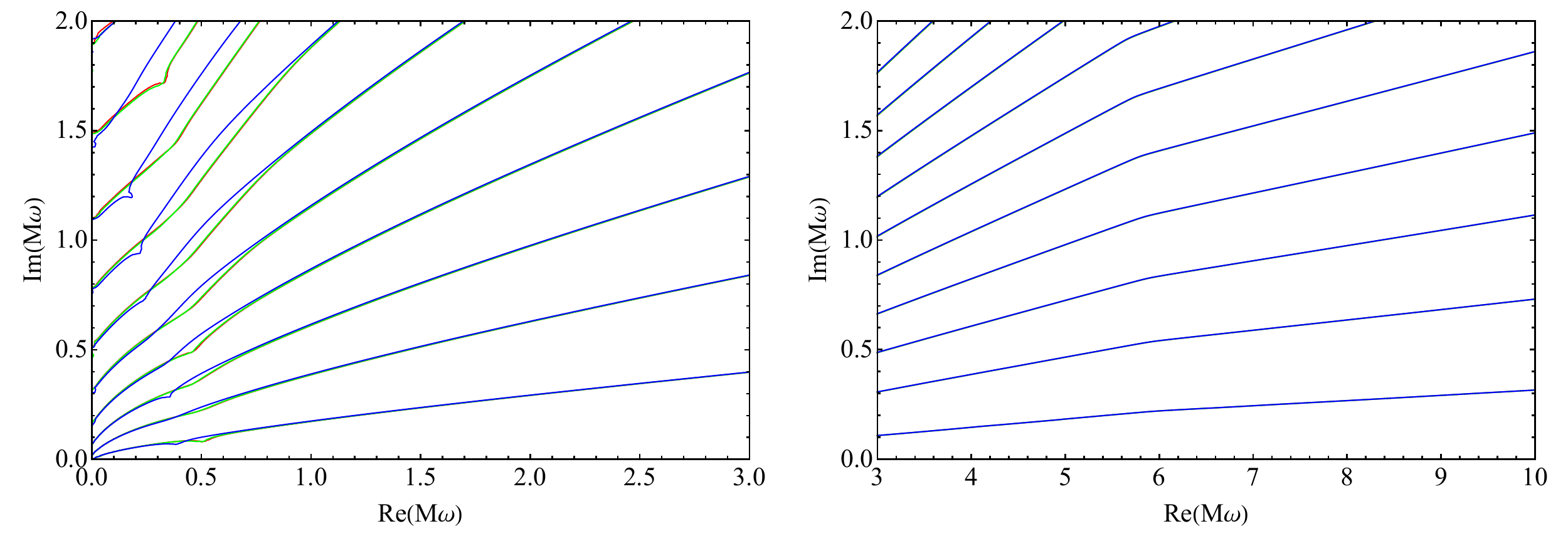}
	\caption{Superimposed scalar (red), electromagnetic-led (green) and gravitational-led (blue) pseudospectral levels for a RN BH with $Q/M=0.5$. All perturbations share the same angular index $\ell=2$. The contour levels $\log_{10}(\epsilon)$ for the left plot range from $-50$ (top level) to $-5$ (bottom level) in steps of $5$; for the right plot, they range from $-20$ (top level) to $-2$ (bottom level) in steps of $2$.}
	\label{asympt}
\end{figure*}

\begin{figure}[t]
	\includegraphics[scale=0.38]{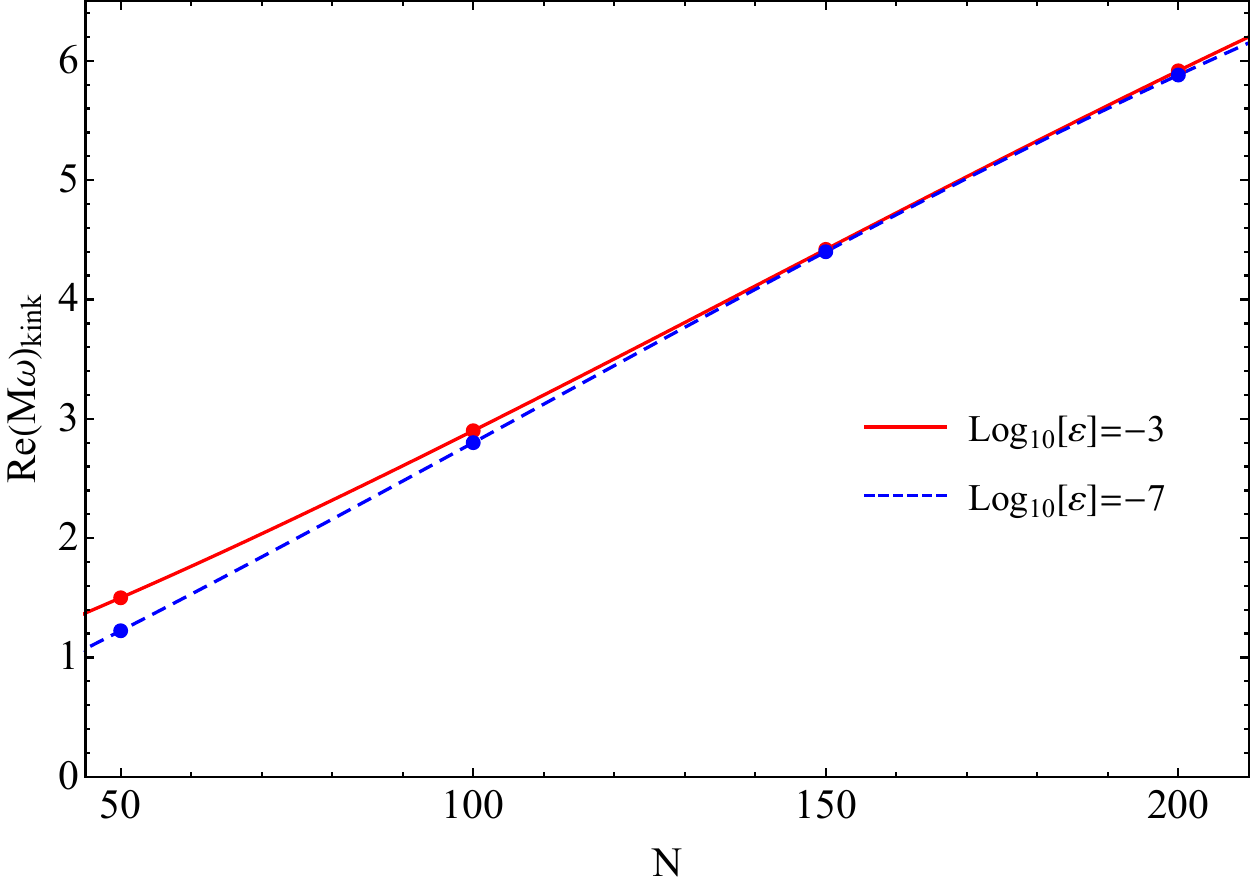}
	\caption{Position on the real axis of the asymptotic kink $\text{Re}(M\omega)_\text{kink}$ for the $\ell=2$ gravitational-led pseudospectral levels of a RN BH with $Q/M=0.5$ with respect to the increment of the grid interpolation points $N$ (see right panel of Fig.~\ref{asympt}).}
	\label{kink}
\end{figure}

We now return to gravitoelectric perturbations. We focus on $\ell-1$ electromagnetic-led and $\ell$ gravitational-led perturbations, which are known to share the same spectra at extremality~\cite{Onozawa:1996ba,Okamura:1997ic,Kallosh:1997ug,Berti:2004md}.

The left panel of Fig.~\ref{grelext} shows the QNM spectra and pseudospectra of $\ell=2$ gravitational-led perturbations, while in the right panel we overplot the pseudospectral contour levels of $\ell=1$ electromagnetic-led and $\ell=2$ gravitational-led perturbations. It is apparent that the isospectrality involved in this case is of a stronger nature. It is not only the spectra that coincide in both cases, but rather the entire pseudospectrum. This occurs because the Green's function as a whole is the same in the two cases.

To understand this claim, recall that the isospectrality between $\ell-1$ electromagnetic-led and $\ell$ gravitational-led fields on an extremal RN geometry is of a different nature from the axial/polar parity isospectrality in the Schwarzschild spacetime~\cite{Chandrasekhar:1985kt}. Here, the isospectrality follows from an invariance of the underlying potential under transformations exchanging the horizon and infinity: see Eq.~(22) in Ref.~\cite{Onozawa:1996ba}, where $r_*\rightarrow -r_*$. 

As discussed in Sec.~\ref{hyperboloidal}, this symmetry is not only restricted to the gravitoelectric potential, but it is a discrete isometry underlying the conformal geometry of extremal BH spacetimes~\cite{Lubbe:2013yia}. In our compactified hyperboloidal coordinates, Eqs.~\eqref{extremal functions} capture the symmetry within the functions in the operator $L$. Comparing the functions present in $L$, the hyperboloidal gravitoelectric potential reads
\beq
q_{(\pm)} = \ell (\ell \mp x) - \left(1 - x^2 \right).
\eeq
for both the $\ell-1$ electromagnetic-led $(-)$ and the $\ell$ gravitational-led $(+)$ fields. 

Thus, the symmetry $x\rightarrow -x$ which maps the horizon to infinity at the level of the conformal geometry is also responsible for mapping electromagnetic-led and gravitational-led potentials.
It then follows that the operators for the gravitational and electromagnetic sectors in the extremal RN spacetime are dual to each other, i.e., 
\beq
L^{(+)}_\ell(x) = L^{(-)}_{\ell -1}(-x).
\eeq
As a consequence, not only do the eigenvalues $\omega^{(+)}_\ell$ and $\omega^{(-)}_{\ell-1}$ coincide, but so do the eigenvectors $v^{(+)}_{\ell}(x) = v^{(-)}_{\ell-1}(-x)$ and the entire pseudospectrum.

\subsection{Asymptotic universality of pseudospectra}\label{asymptotics}

Figure~\ref{asympt} demonstrates the asymptotic structure of three different spectral problems, namely the $\ell=2$ scalar, electromagnetic-led and gravitational-led QNMs of a RN BH with $Q/M=0.5$. The lower pseudospectral levels of the three potentials coincide already at small frequencies, and right beyond the respective QNMs, while higher pseudospectral levels begin to agree with each other only at larger frequencies. Therefore, besides the local region in the non-perturbed QNM vicinity, we find support for an asymptotic universality shared by a whole class of effective potentials, with similar logarithmic patterns. 

An interesting aspect of the asymptotic pseudospectral contours is evident in the right panel of Fig.~\ref{asympt}, where we observe a kink at $\text{Re}(M\omega)\sim 6$. In the range of our analysis, the value of $\text{Re}(M\omega)$ at which the kink occurs decreases as $\log_{10}(\epsilon)$ decreases, slowly moving closer to the imaginary axis. We expect that the kink will eventually meet the imaginary axis for very small $\log_{10}(\epsilon)$. 

Our numerical investigation indicates that this kink does not have a physical origin, but rather is due to the accumulation of numerical error. The curvature sign of the contour changes at the kink, so any fitting beyond the kink itself will not reflect the true logarithmic structure of the pseudospectra. In Fig.~\ref{kink} we verify that the kink is indeed a numerical artifact: as the number of interpolation grid points $N$ increases, the kink moves further away from the imaginary axis. This shows that the well-resolved region in the complex plane grows as we increase the resolution. A similar phenomenon occurs in the calculation of operator eigenvalues through matrix approximations: for a given resolution $N$ only certain eigenvalues can be trusted, but their number increases as $N$ grows. What we observe is the counterpart of this phenomenon at the level of the pseudospectrum, so we expect that the kink should disappear\footnote{Assessing the convergence of the pseudospectra in a quantitative way requires special care, as pointed out in Ref.~\cite{Jaramillo:2020tuu}. The numerical scheme is sensitive to the underlying regularity class of the involved functional spaces. In the limit $N\rightarrow \infty$ one must take into account the regularity issues on the QNM spectra eigenfunctions identified in Refs.~\cite{Ansorg:2016ztf,Gajic:2019oem,Gajic:2019qdd,2020arXiv200407868G}. A rigorous mathematical and numerical convergence analysis is beyond the scope of this work.} 
in the limit $N\rightarrow \infty$.

Taking into account the above limitations of fitting pseudospectral contour lines, we can explore if these branches agree with Eq.~\eqref{e:log_branches}. All branches we have checked in the range $\log_{10}(\epsilon)\in[-3,-35]$ can indeed be fitted accurately by a logarithmic function of the form \eqref{e:log_branches}, in agreement with the Schwarzschild findings in Ref.~\cite{Jaramillo:2021tmt}. (Note in passing that the asymptotics of pseudospectra for the P\"oschl-Teller potential are also described by the logarithmic expression~(\ref{e:log_branches})~\cite{Jaramillo:2020tuu}, but the constants $C_1, C_2$ and $C_3$ have very different values, reflecting the different nature of the P\"oschl-Teller potential at null infinity.) Even though the fits are quite accurate away from the QNM region, where logarithmic asymptotic behavior is expected, Eq.~\eqref{e:log_branches} is a good approximation even close to the QNM corresponding to some given $\epsilon$, in agreement with the discussion in Sec.~\ref{QNM free region}.

\section{Conclusions}

In this work we have presented a detailed study of the pseudospectrum from scalar and gravitoelectric perturbations of the RN spacetime. We observe the same qualitative behavior as in Ref.~\cite{Jaramillo:2020tuu} for all values of the BH charge $Q/M\in[0,1]$ and for all perturbing fields: the pattern of pseudospectral levels is typical of spectrally unstable systems. 

Our numerical calculations reveal a logarithmic dependence of the pseudospectral contour lines, in accordance with theoretical predictions for their asymptotic behavior~(see~\cite{zworski2017mathematical,dyatlov2019mathematical} and references therein). The onset of the logarithmic asymptotic behavior occurs much earlier, in an intermediate frequency regime, and it agrees with the theoretical predictions up to the location of the kink in the pseudospectral contour lines. Note that the kink is a mere numerical artifact, as it moves toward larger values of ${\rm Re}(M\omega)$ as we increase $N$ (i.e., the numerical resolution). These observations confirm the findings of Refs.~\cite{Jaramillo:2020tuu,Jaramillo:2021tmt}: the logarithmic behavior should extend asymptotically to the large-frequency regime.

The hyperboloidal approach to BH perturbation theory plays a crucial role in recasting the underlying wave equation with dissipative boundary conditions into a form best-suited for studying QNMs as the eigenvalue problem of a non-selfadjoint operator. The RN spacetime allows us to examine the effect of different coordinate choices (i.e., different spacetime slicings) on the calculation of the pseudospectra. We have used the so-called areal radius fixing gauge and Cauchy horizon fixing gauges~\cite{PanossoMacedo:2018hab} and found that the behavior of the pseudospectra in the two gauges is nearly identical. This provides strong support to the geometrical nature of BH pseudospectra.

We have paid special attention to the extremal limit $Q/M \rightarrow 1$, characterized by a family of slowly damped modes along the imaginary axis, which approach the value $M\omega=0$ as $Q/M \rightarrow 1$~\cite{Kim:2012mh,Zimmerman:2015trm,Richartz:2015saa,Cardoso:2017soq}. If the underlying potential is slightly modified, a mode arbitrarily close to the real axis is, in principle, prone to cross into the region ${\rm Im}(M\omega) < 0$, and the field's dynamic evolution could display exponentially growing modes.

As observed in hydrodynamics~\cite{Trefethen:1993}, this behavior can be resolved via a pseudospectrum analysis. The pseudospectral contour lines around marginally stable QNMs bound the region where the QNM can migrate under perturbations. If the $\epsilon$-contour line crosses into the unstable region ${\rm Im}(M\omega) < 0$ for small values of $\epsilon$, the likelihood of having an exponentially growing dynamical evolution for a slightly perturbed system is high. We observe such crossings of the pseudospectral contour lines in the lower half of the complex plane for sufficiently small $\epsilon\sim 10^{-4}$, which even agree with the levels considered in the context of turbulence in Ref.~\cite{Trefethen:1993}. However we cannot make any conclusive claims, because we observe a similar behavior even for subextremal RN BHs, presumably due to the presence of (non-convergent) eigenvalues which appear because of our discretization of the differential operators and because of the hyperboloidal compactification in asymptotically flat spacetimes.

Another noteworthy feature of extremal RN BHs is the isospectrality between the QNM frequencies of $\ell-1$ electromagnetic-led and $\ell$ gravitational-led perturbations. We show that the isospectrality is valid not only for the spectrum, but also for the pseudospectrum. This ``strong isospectrality'' is a consequence of a horizon-infinity symmetry which has already been identified as responsible for the gravitoelectric QNM isospectrality~\cite{Onozawa:1996ba}. Such a discrete symmetry is also apparent  in the conformal geometry of extremal BH spacetimes~\cite{Lubbe:2013yia}, and it explains, for instance, the aforementioned transient extremal horizon instability under scalar perturbations~\cite{Aretakis:2011ha,Aretakis:2011hc} in terms of the field properties at future null infinity~\cite{Bizon:2012we,Angelopoulos:2018yvt}. Thus, the horizon-infinity symmetry for the extremal RN BH leads to the Green's functions of electromagnetic-led and gravitational-led perturbations agreeing as a whole, and not just at the poles. 

The plurality of effective potentials describing perturbations of RN BHs allowed us to analyze the asymptotic behavior of pseudospectral contour lines for different spectral problems. By fixing the BH charge and an angular index for scalar, gravitational-led and electromagnetic-led perturbations, we find that beyond the QNM region, $\epsilon$-level sets practically coincide, regardless of the perturbation field. This suggests an ``asymptotic universality'' of pseudospectra. In fact, numerical results for the contour levels are always consistent with a logarithmic asymptotic behavior, in agreement with previous findings for Schwarzschild BHs~\cite{Jaramillo:2021tmt}. 

The pseudospectral analysis presented here can be extended in multiple directions (see also \cite{Jaramillo:2020tuu} for a related list of possible perspectives). It would be interesting to study the superradiant amplification of charged scalar fields~\cite{Bekenstein:1973mi,Denardo:1973pyo}. It is also important to generalize our work to asymptotically de Sitter BHs: the regularity of the cosmological horizon removes the branch cut, and thus the problem becomes closer to the P\"oschl-Teller case studied in \cite{Jaramillo:2020tuu}, which corresponds to the de Sitter spacetime~\cite{Bizon:2020qnd}. By removing the branch cut, the spurious eigenvalues corresponding to the branch cut will also be absent. RN-de Sitter (RNdS) BHs have rich QNM spectra consisting of different mode families~\cite{Cardoso:2017soq}, and as such they provide a perfect testing ground for the pseudospectra of zero modes $M\omega=0$ of neutral scalar fields~\cite{Cardoso:2017soq}, which are prone to superradiant instabilities when the field is charged~\cite{Cardoso:2018nvb,Zhu:2014sya,Konoplya:2014lha,Destounis:2019hca}. Accelerating spacetimes share many similarities with RNdS BHs, with the cosmological horizon being replaced by an acceleration horizon~\cite{Griffiths:2009dfa}. These spacetimes should not be affected by non-convergent ``contaminations'' due to the absence of a non-regular null infinity in the grid, since the boundary conditions are imposed at the acceleration horizon instead~\cite{Hawking:1997ia,Destounis:2020pjk,Destounis:2020yav}.

Another important extension concerns asymptotically anti-de Sitter (AdS) spacetimes. Their timelike null-infinity provides a model for a geometrical ``box'' that suggests interesting analogies with QNM instability problems in optical cavities~\cite{SheJar20}. A study of asymptotically AdS spacetimes requires different boundary conditions. One should similarly introduce different boundary conditions when studying horizonless compact objects, where the BH spectra appear as intermediate time excitations, eventually giving way to ``echoes''~\cite{Cardoso:2016rao,Cardoso:2016oxy,Cardoso:2017cqb,Cardoso:2019rvt}. The relation between pseudospectra, ordinary QNMs and echoes deserves further study.

\section*{Acknowledgments}

K.D. is indebted to George Pappas for very helpful discussions during the early stage of this work. R.P.M. was partially supported by the European Research Council Grant No. ERC-2014-StG 639022-NewNGR ``New frontiers in numerical general relativity''. R.P.M. thanks the warm hospitality of CENTRA-Instituto Superior T\'ecnico (Lisboa) and J.A. Valiente-Kroon for fruitful discussions.
E.B. is supported by NSF Grants No. PHY-1912550 and AST-2006538, NASA ATP Grants No. 17-ATP17-0225 and 19-ATP19-0051, NSF-XSEDE Grant No. PHY-090003, and NSF Grant PHY-20043.
V.C.\ acknowledges financial support provided under the European Union's H2020 ERC 
Consolidator Grant ``Matter and strong-field gravity: New frontiers in Einstein's 
theory'' grant agreement no. MaGRaTh--646597.
J.-L.J thanks the discussions with E. Gasperin, O. Meneses Rojas, L. Al Sheikh and J. Sj\"ostrand, and acknowledges the support of the French ``Investissements d'Avenir'' program through project ISITE-BFC (ANR-15-IDEX-03), the ANR ``Quantum Fields interacting with Geometry'' (QFG) project (ANR-20-CE40-0018-02), the EIPHI Graduate School (ANR-17-EURE-0002) and the Spanish FIS2017-86497-C2-1 project (with FEDER contribution).
This project has received funding from the European Union's Horizon 2020 research and innovation programme under the Marie Sklodowska-Curie grant agreements No 690904 and No 843152.
We thank FCT for financial support through Project~No.~UIDB/00099/2020 and through grants PTDC/MAT-APL/30043/2017 and PTDC/FIS-AST/7002/2020.
The authors would like to acknowledge networking support by the GWverse COST Action 
CA16104, ``Black holes, gravitational waves and fundamental physics.''
Computations were performed on the ``Baltasar Sete-Sois'' cluster at IST, XC40 at YITP in Kyoto University and at Queen Mary's Apocrita HPC facility, supported by QMUL Research-IT.

\appendix

\section{A qualitative picture of spectral stability}\label{stability}

The main goal of our analysis is to compute the pseudospectra of RN BHs, that provide a qualitative understanding of spectral (in)stabilities: how far can points in the spectrum migrate when we perturb the RN operator \eqref{eigenvalue problem}, and what is the associated topographic structure of the pseudospectrum contour lines of the non-perturbed operator? 
\begin{figure}[t!]
	\includegraphics[scale=0.35]{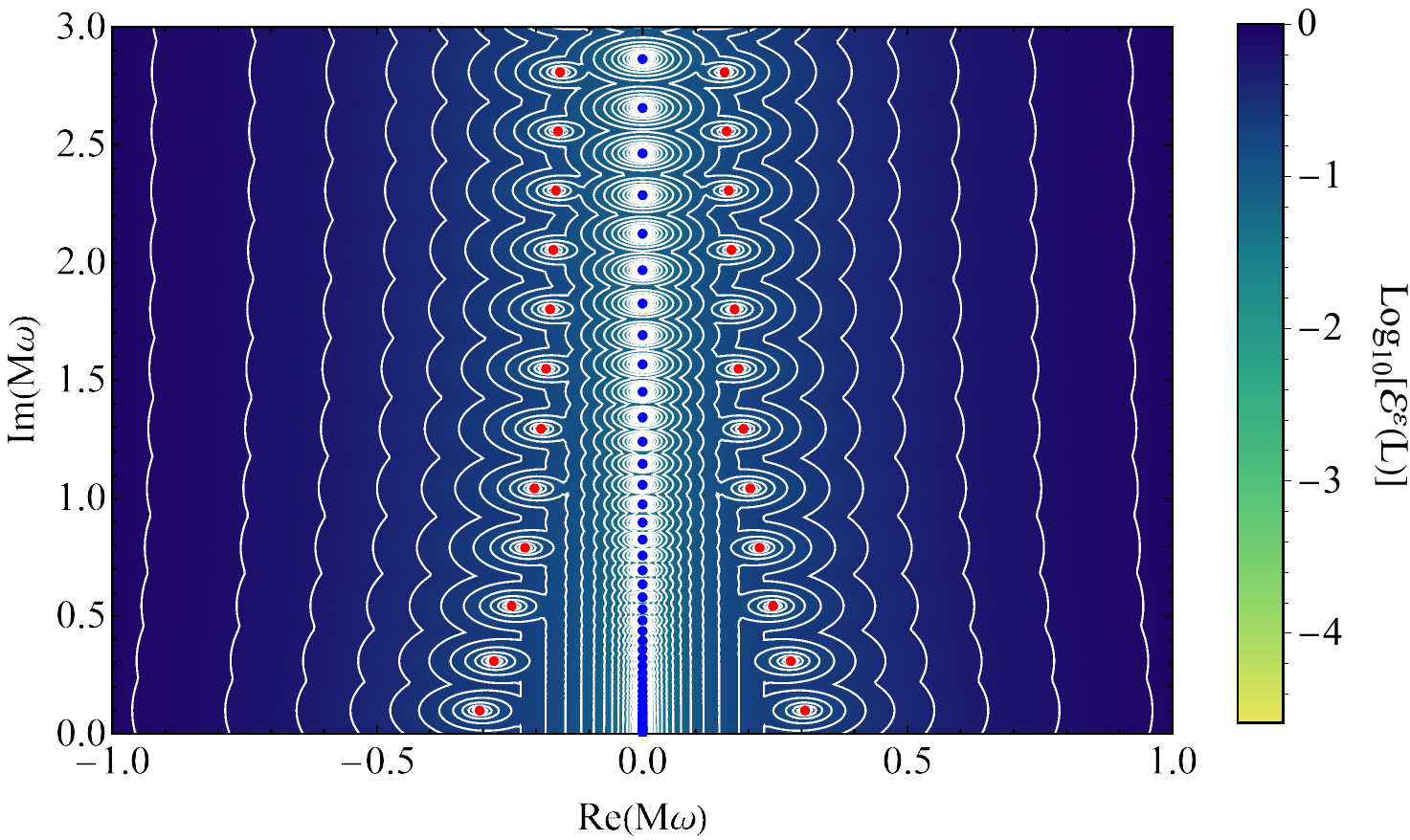}
	\caption{``Error bounds'' $\mathcal{E}^\epsilon(L)$ around the spectrum of $\ell=1$ scalar QNMs of a RN BH with $Q/M=0.5$. The red dots correspond to RN QNMs, while the $\epsilon$-contours illustrate the proximity (to a distance $\epsilon$) of pseudospectra contours to the actual QNM spectrum. The contour levels $\log_{10}(\epsilon)$ range from $-0.04$ (outer) to $-1.94$ (inner) in steps of $0.1$. The blue dots designate branch-cut (non-convergent) modes. This plot shows the typical structure of the pseudospectrum of a spectrally stable (normal) operator.}
	\label{spectral_stability}
\end{figure}

To gain perspective on these questions, in this appendix we consider how a reference case (namely a pseudospectrum exhibiting spectral stability) looks like. The results are easily extended to gravitoelectric perturbations and the qualitative picture remains unchanged.

Let us consider in particular our operator $L$ in Eq. (\ref{matrix evolution}), with $L_1$ and $L_2$ given by expressions (\ref{L1}) and (\ref{L2}) respectively. Using the scalar product (\ref{energy scalar product}), the adjoint $L^\dagger$ is
formally written as~\cite{Jaramillo:2020tuu}
\bea
\label{e:formal_adjoint}
L^\dagger = L + L^{\partial},\ \ L^{\partial} =\frac{1}{i}\!
\left(
\begin{array}{c c}
   0 & 0 \\
   0 & L^\partial_2
\end{array}
\right), 
\eea
where 
\bea
\label{e:L2_boundary}
L^\partial_2 = 2\frac{\gamma(\sigma)}{w(\sigma)}\left(\delta(\sigma)-\delta(\sigma-1)\right).
\eea
The difference $L^{\partial}=L^\dagger - L$ has a special form, with support only on the boundaries ((null infinity at $\sigma=0$ and the horizon at $\sigma=1$), reflecting the fact that the loss of selfadjointness is related to the flux of energy through these boundaries. In fact, as shown in \cite{Gasperin:2021kfv}, the flux of energy at the boundaries depends linearly on the function $\gamma(\sigma)$ at the boundaries.

The previous points suggest the construction of an {\em ad hoc} operator $\tilde{L}$ obtained by setting to zero the value of $\gamma(\sigma)$ at the boundaries, while leaving $\gamma(\sigma)$  in the bulk unaffected. The only purpose of such an ad hoc operator is to illustrate the spectral stability behavior; we emphasize that it does not correspond to any actual evolution operator in a hyperboloidal description of the spacetime. The resulting adjoint $\tilde{L}^\dagger$ is formally identical to $\tilde{L}$. Naively one would expect this to correspond to a selfadjoint operator.
However, the selfadjointness of the differential operator is spoiled, because it also depends on the domain of definition of the operator. The resulting operator is still normal, as we can check by applying Eqs. (\ref{pseudospectra energy norm}) and (\ref{svd definition}) to $\tilde{L}$, with $\tilde{L}^\dagger = \tilde{L}$, leading to the pseudospectrum in Fig.~\ref{spectral_stability} which portrays a normal operator.

Indeed, Fig.~\ref{spectral_stability} captures the overall structure of the would-be pseudospectrum if the actual operator $L$ of scalar perturbations in RN were normal. We observe a clear structure of nested circles with radius $\sim \epsilon$ around the QNMs, while far away from QNMs the contour lines ``flatten out''. This picture is qualitatively identical to a typical normal operator (see Fig.~4 in~\cite{Jaramillo:2020tuu} and \cite{Gasperin:2021kfv}). The non-perturbed eigenvalues of $\tilde{L}$ are the same as the original eigenvalues of $L$, because of the very particular form of the adjoint of $L$. As a consequence, the pseudospectral boundaries of $\tilde{L}$ provide the distances from points in the complex plane to the actual spectrum of $L$, making contact with the $\epsilon$-contour lines $\mathcal{E}^\epsilon(L)$ of the error bound function
introduced  in~\cite{ColRomHan19}.

\bibliography{bibitems}

\end{document}